\documentclass{article}

\usepackage{arxiv}

\usepackage[utf8]{inputenc} % allow utf-8 input
\usepackage[T1]{fontenc}    % use 8-bit T1 fonts
\usepackage{hyperref}       % hyperlinks
\usepackage{url}            % simple URL typesetting
\usepackage{booktabs}       % professional-quality tables
\usepackage{amsfonts}       % blackboard math symbols
\usepackage{nicefrac}       % compact symbols for 1/2, etc.
\usepackage{microtype}      % microtypography
\usepackage{lipsum}		% Can be removed after putting your text content
\usepackage{amsmath}
\usepackage{graphicx}
\usepackage{enumitem}
\usepackage{tabularx}
\usepackage{comment}
\usepackage{subcaption}
\usepackage{multirow}
\usepackage{array, caption, boldline}
\usepackage{cellspace}
\usepackage{hhline}

\usepackage{float}
\floatstyle{plaintop}
\restylefloat{table}

\title{Discerning Reliable Cyber Threat Indicators for Timely Cyber Threat Intelligence}

\author{Dincy R Arikkat, Vinod P.\thanks{Department of Mathematics,  University of Padua, Italy , vinod.p@math.unipd.it}, Rafidha Rehiman K. A.\\
	Department of Computer Applications\\
 Cochin University of Science and Technology\\
 Kochi, India \\
	\texttt{\{dincyrarikkat,vinod.p,rafidharehimanka\}@cusat.ac.in} 
	\And
	Andrea Di Sorbo, Corrado A. Visaggio\\
	Department of Engineering\\
 University of Sannio\\
 Benevento, Italy\\
	\texttt{\{disorbo,visaggio\}@unisannio.it} 
 \And
	Mauro Conti \\
	Department of Mathematics\\
 University of Padua, Italy \\
	\texttt{mauro.conti@math.unipd.it} 
}

\date{}

\hypersetup{
pdftitle={Can Twitter be Used to Acquire Reliable Alerts against Novel Cyber Attacks?},
pdfsubject={Cyber Threat Intelligence},
pdfauthor={Dincy R Arikkat,Vinod P, Rafidha Rehiman K A, Andrea Di Sorbo, Corrado A. Visaggio, Mauro Conti},
pdfkeywords={Cyber Threat Intelligence~(CTI),Indicators of Compromise~(IoC),Open Source Intelligence~(OSINT),Threat Hunting,Threat Intelligence}
}

\begin{document}
\maketitle

\begin{abstract}
	In today's dynamic cybersecurity landscape, timely and accurate threat intelligence is essential for proactive defense. This study explores the potential of social media platforms as a valuable resource for extracting actionable Indicators of Compromise (IoCs). Utilizing a Convolutional Neural Network (CNN), we achieved an F1-score of 98.80\% and a detection rate of 99.65\%, filtering vast social media data to identify key IoCs, including IP addresses, URLs, file hashes, domain addresses, and CVE IDs. These indicators are critical for detecting potential threats and vulnerabilities, and their relevance was evaluated using metrics such as \textit{correctness}, \textit{timeliness}, and \textit{overlap}. Our analysis shows that URLs emerged as the most frequently shared IoC, with 48.67\% representing valid threats. To further investigate the role of automated accounts in disseminating IoCs, we applied several machine learning models, with XGBoost delivering the highest performance—achieving a macro F1-score of 0.814 and a weighted F1-score of 0.925. These findings highlight the growing significance of social media as a reliable source of actionable threat intelligence, offering valuable insights for cybersecurity professionals to stay ahead of emerging threats.
\end{abstract}

% keywords can be removed

\section{Introduction}
\label{sec:intro}
The digital revolution has led to remarkable progress, including reduced operating costs, improved efficiency, faster communication, and enhanced system accessibility, but it has also introduced significant cybersecurity challenges~\cite{hasan2021evaluating}. As organizations increasingly rely on interconnected systems and digital platforms to conduct business, they become more vulnerable to cyber-attacks, ranging from data breaches to ransomware attacks\footnote{https://www.kaspersky.co.in/blog/ransowmare-attacks-in-2023/27107/}. These threats can have catastrophic consequences, compromising the integrity, confidentiality, and availability of sensitive information and critical infrastructure~\cite{lone2023comprehensive}. Also, the emergence of different Advanced Persistent Threat~(APT) groups, such as Snake, Volt Typhoon, Scattered Spider, etc, further exacerbates these challenges. These groups leverage different Tactics, Techniques, and Procedures~(TTPs) to target high-value assets for espionage, sabotage, or financial gain\footnote{https://www.ibm.com/reports/data-breach}. According to Checkpoint~\footnote{https://blog.checkpoint.com/research/check-point-research-2023-the-year-of-mega-ransomware-attacks-with-unprecedented-impact-on-global-organizations/} analysis, in 2023, the cyber threat landscape witnessed a continuous surge in attacks, through which organizations encountering an average of 1,158 weekly cyber assaults. Notably, there was a significant shift in ransomware tactics, with attackers emphasizing data theft and extortion-based campaigns over traditional encryption-based ransomware. Specifically, campaigns like MOVEit and GoAnywhere exemplify this strategic shift in cyberattacks. So, understanding attacker’s methods and infrastructure necessitates comprehensive threat intelligence derived from various sources, including social media, hacker forums, technical bulletins, and the dark web, to effectively detect and thwart their activities~\cite{zhao2020timiner}~\cite{bromiley2016threat}.
\par
Furthermore, the recent attacks and vulnerabilities highlight the critical need for such threat intelligence. For instance, the T-Mobile data breach~\footnote{https://www.bleepingcomputer.com/news/security/t-mobile-hacked-to-steal-data-of-37-million-accounts-in-api-data-breach/} in January 2023 resulted in the theft of personal data from over 37 million customer accounts. In this incident, attackers exploited a vulnerability in a supplier's Application Programming Interface~(API) to gain unauthorized access to the personal data of these customers. %The T-Mobile attack underscores the critical importance of safeguarding personal data. 
Also, the emergence of vulnerabilities like CVE-2023-43208 and CVE-2024-236191, impacting the healthcare sector, and CVE-2024-21736, affecting finance, underscores the vital importance of gathering proactive threat intelligence. Such types of vulnerabilities have led to widespread exploitation and data breaches across numerous organizations globally. These incidents emphasize the need for Cyber Threat Intelligence~(CTI) to identify and mitigate such vulnerabilities before they can be exploited by malicious actors. By aggregating, analyzing, and disseminating actionable insights from diverse sources, CTI empowers security professionals to stay ahead of evolving threats~\cite{wagner2019cyber}. One crucial aspect of CTI is the collection of Indicators of Compromises (IoCs), which serve as tangible artifacts indicative of malicious activity~\cite{asiri2023understanding}. They are observable evidence of compromises that can help to identify the impact of the attack and understand how to cleanse the affected perimeter and restore the operation of affected networks, devices, and services. IoCs encompass IP addresses, URLs, file hashes, domain addresses, and Common Vulnerabilities and Exposures~(CVE) IDs. 
\par
CTIs are traditionally derived from official sources, including the National Vulnerability Database~(NVD) and the database, yet there are substantial gaps between vulnerability notification and NVD publication~\cite{alves2020follow}. These methods struggle to match the real-time pace of social media. This is where Open-Source Threat Intelligence~(OSINT) techniques become crucial. OSINT offers early warnings of events, vulnerabilities, and exploits due to the instantaneous nature of social media information dissemination. Social media platforms serve as fertile grounds for reconnaissance and information gathering, providing invaluable resources for identifying early indicators of cyber threats. Recent studies highlight that security experts frequently utilize social media channels to distribute IoCs\cite{shin2021twiti,alves2020follow}.
\par
However, the sheer volume of data generated on social media poses significant challenges for threat analysts. The manual analysis of vast amounts of unstructured social media data is time-consuming, labor-intensive, and prone to errors. To address these challenges, we employed techniques such as Regular expression and Machine Learning~(ML) to filter valuable threat indicators for further investigation. We comprehensively analyze IoCs from social media platforms with diverse threat intelligence services and evaluate how precise such IoCs are and how timely intelligence they provide. Our findings show that 98.08\% of the file hashes mentioned in social media are reliable IoCs, whereas social media delivers 1.27\% of novel vulnerabilities that are not present in the NVD database. Moreover, 46.98\% of URLs, 28.75\% of IP addresses, and 56.77\% of domain names indicated in social media posts are trustworthy cyber forensic artifacts.
\par
Furthermore, the prevalence of automated social media accounts necessitates examining their role in disseminating IoCs. To achieve this, we employed Machine Learning (ML) and Deep Learning (DL) models to develop an automated system for distinguishing between human-operated and automated accounts sharing IoCs. Also, we incorporated Explainable Artificial Intelligence (XAI) techniques to understand the rationale behind the system's decisions. Finally, we assessed the proportion of automated accounts involved in disseminating information related to security threats. Our analysis confirms that automated accounts are also involved and play a complementary role in disseminating confirmed IoCs on social media.
\par
The major contributions of this paper are summarized as follows:
\begin{itemize}
    \item To facilitate automatic threat intelligence gathering from social media platforms, we devised an automated approach for harvesting and extracting Indicators of Compromise (IoCs) such as IP addresses, URLs, domain addresses, file hashes, and CVEs. This method aims to determine the extent to which social media platforms offer timely and varied threat intelligence regarding security vulnerabilities.
    \item We conducted an analysis of the reliability of IoCs shared across social media posts to enhance threat detection and response strategies within the cybersecurity sphere. Using threat intelligence services such as VirusTotal, AlienVault, Urlhaus, MalwareBazaar, MISP, and the CVE database, we evaluated the accuracy, timeliness, and overlap of these IoCs.
    \item We assembled a comprehensive dataset\footnote{The dataset is available at \url{https://github.com/OPTIMA-CTI/Twitter-CTI.git}} containing a multitude of features, including account-level attributes, post-related information, and temporal data. This dataset facilitates the classification of social media accounts into categories of human-operated or automated. It serves as a valuable resource for training and assessing machine learning models designed to identify automated accounts engaged in the dissemination of threat indicators.
    \item We implemented a generic system to identify automatic accounts sharing threat indicators using Machine Learning and Deep Learning models. Furthermore, the model is interpreted using various explainability tools to identify the important features involved in the decision process.
\end{itemize}

The rest of the paper is organized as follows. Section~\ref{sec:Related} discusses related works. The objectives and proposed methodology are elaborated in Section~\ref{sec:design}. Section~\ref{sec:security_tweets} discusses data collection strategy and identification of relevant security posts. Extraction and analysis of IoCs are described in Section~\ref{sec:ioc_extraction}. Section~\ref{botdataset} gives a detailed description of features generated for identifying automated accounts. The implementation and interpretation of the bot classification model are discussed in Section~\ref{bot_classification}. Then, in Section~\ref{sec:results}, we present findings about the research questions, and Section~\ref{sec:discussion} elaborates discussions. Finally, Section~\ref{sec:conclusions} draws the conclusions and future scope.
\section{Related Work}
\label{sec:Related}
In this section, we discuss the related literature about IoCs extraction, IoCs validation, and profile validation. 
\subsection{IoCs Extraction.} Nowadays, social media can aid researchers in conducting threat hunting to generate CTI. Most of the state of the art tried to propose various techniques for extracting CTI data like network IoCs~(\textit{e.g.,} URL, IP, and hash address) from threat reports, online articles, and social media. Some research works rely on static CTI extraction. The work done by Liao \textit{et al.,} \cite{liao2016acing} proposed a tool for extracting IoCs from 71,000 unstructured articles of 45 technical blogs. They utilized Natural Language Processing~(NLP) techniques such as Relation Extraction~(RE) and Named Entity Recognition~(NER) to extract IoCs like IP, hash, and DateTime. Zhu \textit{et al.}\cite{zhu2018chainsmith} proposed an approach to extract URLs and IP addresses from security articles by leveraging the semantics of malicious campaigns found in threat intelligence reports. In extracting IoCs, their study achieved a precision and recall rate of 91.9 and 97.8, respectively. Furthermore, few researchers focused on descriptive CTI extraction that not only extracts static indicators but also understands the nature of the threats in detail~(\textit{e.g., attack vectors, tool)}. For instance, Husari \textit{et al.}~\cite{husari2017ttpdrill} implemented a tool, named  TTPDrill, for extracting threat actions from CTI reports using NLP and Information Retrieval techniques. They also  created a comprehensive attack model by linking each identified threat action to the corresponding techniques, tactics, and kill chain phases. TTPDrill obtained a precision and recall rate of 84\% and 82\%, respectively. Zhao \textit{et al.}\cite{zhao2020timiner} extracted IoCs from blogs and news articles and categorized these CTI-related contents based on (i) domain tags like phishing or malware.
\par
Long \textit{et al.} \cite{long2019collecting}, applied Neural-Based sequence labeling on English and Chinese cybersecurity articles for extracting IoCs. Their model has achieved an average F1-score of 89.0\% for English and 81.8\% for Chinese articles. Researchers also investigated IoCs over social networks such as Twitter and Facebook. Niakanlahiji \textit{et al.}~\cite{9006562} conducted experiments on Twitter posts and proposed a system that incorporates graph theory to identify credible Twitter users. They also proposed machine learning algorithms to distinguish CTI tweets from non-CTI tweets and text mining methods to extract IoCs. The study observed that around 10\% of identified URLs exist in blacklisted databases when considering real-time extraction, and it increased to 26\% after one week. In~\cite{sabottke2015vulnerability}, Sabottke \textit{et al.}  gathered data by searching tweets with  CVE identifiers. Since tweets without the CVE ID are not considered, the system is insufficient to discover possible exploits of other threat-related tweets. 
\subsection{IoCs Validation.} As the cyber threat evolves at a rapid rate, security specialists face various challenges when analyzing information that draws from various sources. Multiple steps need to be undertaken to exchange information related to potential cyber threats, which must be completed before the information can be considered actionable CTI. Several sources have discussed threat intelligence's actionability and observed various traits that contribute to deciding its usefulness. These characteristics include trustworthiness, timeliness, anonymity, relevance, reputation, and the ability to exchange data seamlessly between different systems~\cite {wagner2019cyber}. Reliability of CTI focused on two aspects (i) the CTI source and (ii) the data provided by these sources. In either case, evaluating the trustworthiness of a CTI transmitted by a source is necessary. Schaberreiter \textit{et al.}~\cite{schaberreiter2019quantitative} proposed a set of parameters such as similarity, completeness, timeliness, compliance, interoperability, verifiability, false positives, maintenance, and extensiveness to evaluate a CTI source and introduced a trust measure. In another work, Ibrahim \textit{et al.}~\cite{al2017beyond} investigated the indicator's quality  for assessing the level of contribution made by participants in sharing information for threat intelligence. They evaluated the quality of the indicators by considering factors such as uniqueness, correctness, utility, and relevance. 
\par
Researchers in~\cite{ermerins2020scoring} presented an IoC scoring model to diminish the amount of false positives by combining intelligence feeds. They collected the IoCs from several threat intelligence platforms like Cyber Cure, C\&C Tracker, Binary Defense, and AbuseIPDB. They computed the extensiveness, timeliness, completeness, and whitelist overlap for each feed and measured the source confidence with each feed. Alves et al.~\cite{alves2020follow} compared multiple vulnerability repositories and Twitter to study the latency with which vulnerabilities were published on various platforms. They found that Twitter generated early security warnings compared to the NVD database. Shin \textit{et al.}~\cite{shin2021twiti} developed a tool for extracting malware IoCs such as IP, Domain, URL, and file hash within Twitter posts. They showed that the performance of Twitter for capturing some malware families like Emotet is better than the other threat intelligence feeds. For each IoC type, they also evaluated the accuracy, latency, and exclusiveness of IoCs against different Threat Intelligent Services such as AlienVault OTX, MalwareBazaar, and Virus Total. They came to the conclusion that extracted IoCs are unique and early.
\subsection{Profile Validation.} 
Several researchers have investigated how to determine the authenticity of social media accounts. Adewole \textit{et al.}~\cite{adewole2019smsad} proposed a model for detecting spam messages and accounts in Online Social Networking~(OSN). To identify spam messages, they compiled a dataset from three sources: Twitter Spam Corpus, SMS Corpus V.0.1 Big, and SMS collection V.1, which contained 18,000, 1324, 5574 and  samples, respectively. Their model relied on 18 features, including content and behavior-based characteristics, and the RF classifier obtained a precision of 0.933 and an area under the curve~(AUC) of 0.977. Alom \textit{et al.}~\cite{alom2020deep} proposed a deep learning strategy for identifying Twitter spammers based on tweets and meta-data such as account age, number of replies, followers/friends, etc., to understand different characteristics affecting the classification. DABot~\cite{wu2021novel} uses a Residual Network~(ResNet), an attention mechanism, and a bidirectional Gated Recurrent unit~(BiGRU) to recognize social bots and legitimate users. It identifies social bots from humans based on a variety of characteristics, including user-level, interaction-based, time-related, and content-related characteristics. 
\par
Bose \textit{et al.}~\cite{bose2021tracing} developed a semi-supervised technique using a Regression model based on user weight calculation, community formation, and central user impact assessment to trace and promote Twitter user profiles that can contribute as instances of CTI-related information based on a collection of supplied seed nodes. Aswani \textit{et al.}~\cite{aswani2018detection} developed a system for detecting Twitter spammers using a bio-inspired algorithm called Firefly. Their system achieved a high accuracy rate of 97.98\% and relied on features such as unique words, diversity of hashtag frequency, polarity, etc. Heidari \textit{et al.}~\cite{heidari2021empirical} proposed a bot detection model by extracting sentiment features from tweet texts. Their study demonstrated the effectiveness of this approach on Dutch tweets and achieved an accuracy of more than 87\%. Rodr{\'\i}guez-Ruiz \textit{et al.}~\cite{rodriguez2020one} implemented a one-class classification for better Twitter bot detection. Their experiment showed that this method had a consistent performance in detecting different types of bots, achieving an AUC score above 0.89 without prior knowledge about the bots.
\par
While several studies have explored IoC extraction from blog articles and social media, few of them have investigated the reliability of these indicators. To advance this field, our investigation delves into the accuracy, timeliness, and overlap of IoCs transmitted through social media, utilizing popular Threat Intelligence platforms for validation. Additionally, we surpass the current state of the art by examining the prevalence of IoCs common to all platforms to determine the most popular indicators. Notably, no prior research has examined the role of automated accounts in delivering IoCs. Therefore, our study aims to fill this gap by exploring the involvement of these accounts in IoC dissemination.
\section{Study Design}
\label{sec:design}
In this section, we discuss the objectives and context of the study, the research questions, and the architecture of the system conceived to answer them.
\subsection{Objectives and Research Questions}
This investigation aims to understand to which extent social media can provide data to derive actionable threat intelligence and take appropriate actions to subvert cyber-attacks.
The study focuses on three goals: (i) to obtain a characterization of the threat intelligence (\textit{i.e.,} IoCs) delivered with social media, (ii) to assess the reliability of the IoCs distributed with security-related social media profiles, and (iii) to develop a model for determining the involvement of the automatic accounts in disseminating IoCs. Our three goals have been declined in three research questions as below: \\ 
\textbf{\textit{RQ$_{1}$}: Which kinds of IoCs are mainly transmitted through social media?}\\
	To provide early alert against novel attacks, a multitude of indicators of compromise is being distributed through social media, including Twitter, blog posts, discussion forums, and other similar platforms. Each of these sources shares different amounts and types of IoCs, among which we can refer to email indicators, host-based indicators, network indicators, and behavioral indicators \footnote{\url{https://resources.infosecinstitute.com/topic/threat-hunting-iocs-and-artifacts/}}. In this paper, we are focusing on network indicators (such as IP, URL, domain addresses), file hashes, and vulnerability indicators such as CVE ID to measure the relevance of social media in CTI. Therefore, with this question, we aim to monitor the kind of network and vulnerability IoCs commonly shared via social media posts. \\ 
 \textbf{\textit{RQ$_2$}: To what extent IoCs extracted from social media accounts are reliable?}\\
	Analyzing the trustworthiness of IoCs derived from social media posts is a key step toward understanding how we can leverage social media to identify novel attack patterns. Information from Threat Intelligence Services~(TIS) is used to validate the reliability of IoCs through diverse performance indicators and to answer specific research questions. To evaluate the significance of social media in CTI, we, in particular, evaluate three metrics and describe them based on the following research questions. \\
       \textbf{\textit{RQ$_{2.1}$} - Correctness}: \textit{What percentage of IoCs are reported malicious by different threat intelligence services?}\\
        Correctness determines whether the IoCs mentioned in the social media posts are also mentioned by other threat intelligence services. We considered an IoC as confirmed~(threat indicator) if any of the TIS reported it as malicious. Equation~\ref{eq:correctness} is used to compute correctness:
        \begin{equation}
            \mathrm{Correctness}=\left(\frac{nMalIoC}{nIoC}\right)*100
            \label{eq:correctness}
        \end{equation}
    where $nIoC$ represents the total number of IoCs gathered from the social media posts and $nMalIoC$ is the number of IoCs found on social media which are also found on other TIS.\\
	 \textbf{\textit{RQ$_{2.2}$} - Timeliness}: \textit{How fast does social media share its IoCs compared to other threat intelligence services?}\\
	 This measure indicates which sources are usually early in providing threat intelligence relating to the same event. To quantify such measure, we use the following equation:
    \begin{equation}
        \Delta{\mathcal{T}} = \mathcal{T}_{twitter}- \mathcal{T}_{tis}
    \end{equation}    
    where $\mathcal{T}_{twitter}$ is the reported date of IoC in security related social media posts and $\mathcal{T}_{tis}$ is the start date of the same IoC in the other threat intelligence services.\\
    \textbf{\textit{RQ$_{2.3}$} - Overlap}: \textit{How many IoCs in social media exist in other Threat Intelligence Services?}\\
    The existence of a single IoC over different TISs indicates its relevancy. This criterion motivates if the information associated with a given IoC is useful to identify a threat. This attribute is calculated by determining the number of IoCs common in different TIS.\\
 \textbf{\textit{RQ$_3$:} To what proportion are the automated accounts delivering IoCs through social media?}\\
On social media, different accounts can provide threat-related information, including legitimate accounts managed by human security experts and automated accounts that rely on bots or scripts. Legitimate accounts may provide a higher proportion of verified and high-quality threat-related data, as they can often perform more nuanced analysis and verification of IoCs before sharing them. On the other hand, automated accounts can be beneficial for rapidly detecting and sharing IoCs with security teams and may provide a higher volume of threat-related data overall. However, an automated account may disseminate misleading threat intelligence data and propagate false flags~\cite{ranade2021generating}. This research question determines the proportion of legitimate and automated accounts disclosing IoCs on social media. Equation~\ref{eq:proportion} calculates the proportion of automated accounts delivering cyber forensic artifacts.
\begin{equation}
        \mathrm{PropBot}=\left(\frac{nBotMalIoC}{nTotMalIoC}\right)*100
        \label{eq:proportion}
    \end{equation}
     where $nBotMalIoC$ represents the total number of malicious IoCs reported by automated accounts, and $nTotMalIoC$ is the total number of malicious IoCs determined using TIS.
\subsection{Context Selection}
In this section, we describe the context of our investigation. For this study, we focus on Twitter~(now known as X) as a social media platform from which we collected data. We also provide background information about IoCs, and the Threat Intelligence Services.
\subsubsection{Twitter}
Twitter(X) is a popular online microblogging service, and it is also used for discussing news related to different topics, like generic information, politics, economics, entertainment, and technology~\cite{shin2020cybersecurity}. Furthermore, Twitter has 396.5 million accounts and 206 million users accessing their accounts daily\footnote{https://www.brandwatch.com/blog/twitter-stats-and-statistics/}. Due to the nature of timeliness, diversity, and volume, Twitter seems to be one of the prominent sources for deriving threat intelligence compared to other social media platforms~\cite{yagcioglu2019detecting}. In~\cite{horawalavithana2019mentions}, Horawalavithana \textit{et al.} reported that many vulnerabilities are disclosed on Twitter. Recently, many security practitioners have used Twitter to disseminate their findings of threats in the form of IoCs, as well as discuss data breaches, zero-day vulnerabilities, DDoS, and ransomware attacks\cite{shin2020new}~\cite{sabottke2015vulnerability}~\cite{altalhi2021survey}. For instance, in June 2017, an incident employing \textit{NotPetya} was mentioned on Twitter prior to getting published by a mainstream source~\cite{sapienza2018discover}. 
\subsubsection{Indicators of Compromise} Indicators of compromise are observable computer forensic artifacts that are specific to a threat, and that can be used to detect an attack and locate the targeted machines. IoCs are usually extracted from OSINT platforms and reveal information about the exploitation and capabilities of adversaries. Network, host-based, and email IoCs are the three predominant types of IoCs that provide information for cyber forensic investigations and give an indication of the attack\footnote{https://resources.infosecinstitute.com/topics/threat-hunting/threat-hunting-iocs-and-artifacts/}. Network IoCs, such as URLs and domain names, can be related to Command and Controls or drop servers used in botnets or generally in the communication of malware. Additionally, IP addresses aid in identifying attacks emanating from a known corrupted server; however, IP indicators may last a limited time as attackers are used to frequently moving their servers to other locations. The host-based IoCs comprise hashes of the file, the mutex, the registry keys, and the Dynamic Link Libraries~(DLLs). On social media sites, file hashes like MD5 and SHA256 are commonly reported indicators used to recognize the malicious program uniquely, but recently imphash and SHA3-384 have been added. Nowadays, the distribution of malware is spread through email links as well as email attachments, so this forms another type of IoC.
\par Furthermore, recent studies demonstrated how the detection of advanced persistent threats is made simpler by vulnerability indicators like CVE-ID, a standard index form for known vulnerabilities~\cite{alves2020follow}. A short overview of the vulnerability, along with the release date, can be discovered by the CVE-ID. Hence, gathering and extracting relevant IoCs from sources like articles, hacker forums, blogs, and social media posts, leads to deriving useful CTI that will allow organizations to prepare ahead for attack campaigns. 
\subsubsection{Threat Intelligence Services} Several Threat Intelligence Services~(TISs) are available to gather and organize threat information and IoC. The TIS gathers information from diverse sources on a regular basis to help security analysts investigate threats and provide early warnings about attacks. Nowadays, there are several TIS available online and freely. Researchers utilize VirusTotal to evaluate the files and gather threat information~\cite{peng2019opening}. VirusTotal~\footnote{https://support.virustotal.com/hc/en-us/articles/115002126889-How-it-works} is one of the reliable TIS, employing 75 antivirus software and 62 domain and website scan engines. Further, AlienVault OTX\footnote{https://otx.alienvault.com/} can be freely accessible to security experts and threat hunters to enable collaborative investigation. Security experts can add suspicious or malicious threat information in the form of a pulse on the AlienVault platform. Both VirusTotal and AlienVault communicate with the server via a unique API key. URLhaus\footnote{https://urlhaus.abuse.ch/} is a project of abuse.ch that captures, monitors, and distributes malicious URLs to assist security experts to protect their systems from cyber-attacks. URLhaus does not require an API key to connect with the server; direct communication can be established. MalwareBazaar\footnote{https://bazaar.abuse.ch/}, like URLhaus, is an abuse.ch project and does not require an API connection to examine IoCs in its database. MalwareBazaar accepts hashes rather than IP, URL, domain, and CVEs compared to other threat intelligence services. The Malware Information Sharing Platform~(MISP) is an open-source application that enables the collection and sharing of IoCs along with vulnerability information~\cite{wagner2016misp}. In MISP, threat information is reported in the form of events: contextual description, threat level, actors, date, etc. All forms of IoCs, including CVE-ID, are searchable in MISP. On the other hand, the National Vulnerability Database~(NVD) focuses primarily on collecting information related to software vulnerabilities. As NVD and the CVE List are fully synced, any revisions to CVE are instantly reflected in NVD. The NIST undertakes the continuing investigation of CVEs and determines threat severity for each vulnerability using Common Vulnerability Scoring System~(CVSS)~\cite{booth2013national}. Further, sandbox environments such as JoeSandbox\footnote{https://www.joesandbox.com/\#windows}, Any.run\footnote{https://any.run/}, and Intezer\footnote{https://www.intezer.com/} provide detailed information about malware from which IoCs can be extracted. However, in this study, we employ Threat Intelligence Platforms such as VirusTotal, AlienVault, UrlHaus, MalwareBazaar, MISP, and vulnerability databases like NVD, which were selected based on  existing literature~\cite{nayak2019analyzing}~\cite{shin2021twiti}~\cite{liao2016acing}~\cite{wagner2016misp}~\cite{dong2019towards} and their prevalent usage. The acceptable levels of various indicators of compromise in TISs are summarized in Table~\ref{tab:TIP_Input}
\begin{table}[h!]
\caption{IoC reported by Threat Intelligence Services}
\tiny
% Row extra space
\setlength{\extrarowheight}{3pt}
\centering
    
   \begin{tabular}{llllll}
   \hline
      \textbf{TIS}& \textbf{URL} & \textbf{IP} & \textbf{Domain}& \textbf{Hash} & \textbf{CVE}\\
      \hline
      \textbf{VirusTotal} & \checkmark & \checkmark & \checkmark &	\checkmark & \textbf{X}\\

      \textbf{AlienVault} & \checkmark & \checkmark & \checkmark &	\checkmark & \textbf{X}\\
 
      \textbf{URLhaus} & \checkmark & \checkmark & \checkmark &	\checkmark & \textbf{X}\\
 
      \textbf{MalwareBazaar} &\textbf{X}  & \textbf{X}  &\textbf{X}   & \checkmark &	\textbf{X}\\

      \textbf{MISP} &\textbf{X}  & \textbf{X}  &\textbf{X}   & \checkmark &	\checkmark\\

      \textbf{NVD} &\textbf{X}  & \textbf{X}  &\textbf{X}   & \textbf{X} &	\checkmark\\
      \hline
    \end{tabular}
  \label{tab:TIP_Input}
\end{table} 
\subsection{System Architecture}
\begin{figure*}[ht]
    \includegraphics[width=\textwidth]{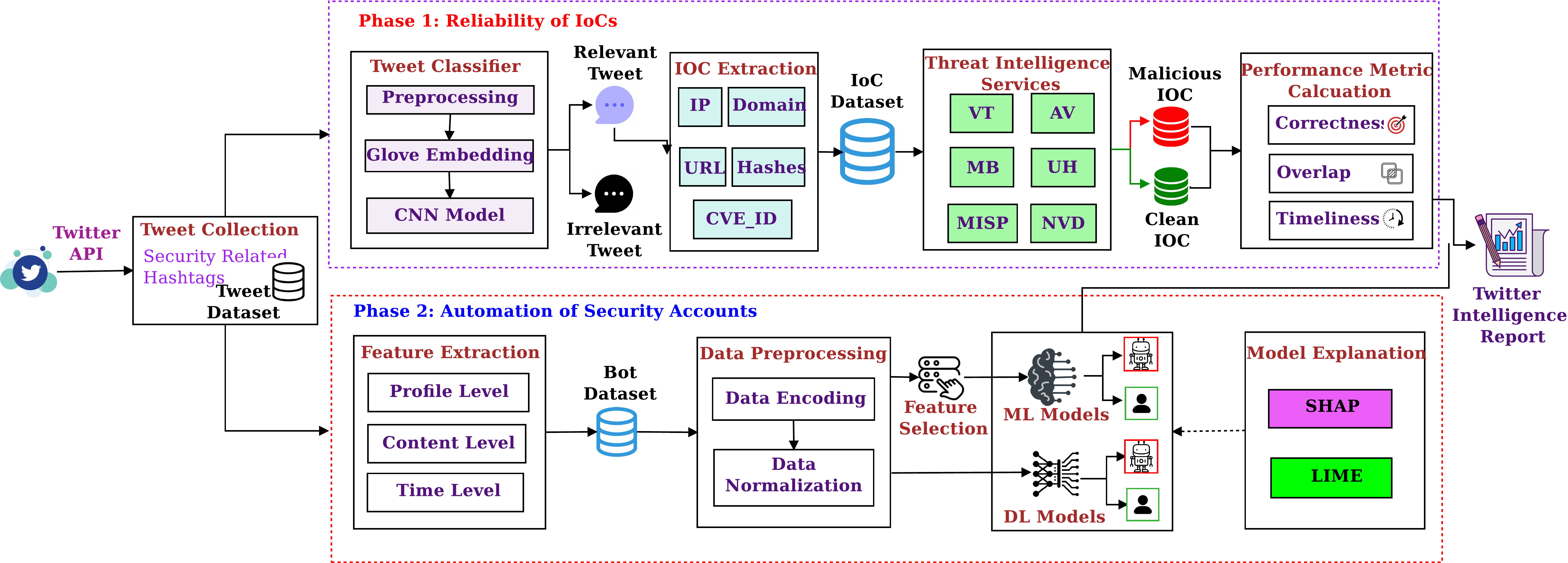}
    \caption{System Architecture of Threat Intelligence from Social Media}
    \label{fig:architecture}
\end{figure*}

Figure~\ref{fig:architecture} depicts the proposed system for extracting Threat Intelligence from Twitter. The methodology involves two phases namely: Reliability of IoCs~(RoIoC) and Automation of Security Accounts~(AoSA).
In the \textit{RoIoC phase}, we evaluate the reliability of IoCs and analyze how quickly Twitter distributes threat information aids in formulating actionable threat intelligence. In the initial process, we gathered seven months of security tweets published from January 2021 to June 2021. The process for collecting the tweets is described in Section~\ref{sec:security_tweets}. While we acquired Twitter data as an IoC source for our study, it is possible that we also collected non-security tweets or tweets without IoCs. In our study, we considered a relevant tweet as one that includes an IoC. However, some tweets may be malware tutorial tweets, references to security blogs, or includes URLs such as Facebook, X~(Twitter), and Reddit links that can not be considered valid IoCs. So, we regarded tweets without IoCs as irrelevant tweets. To exclude irrelevant tweets, we develop a filtering system based on NLP and DL models. After identifying relevant tweets, we extracted network IoCs and vulnerability IoCs. We employed various TISs to extract additional information related to IoCs and determined the threat metrics such as \textit{correctness, timeliness}, and \textit{overlap} to estimate the reliability of IoC. Subsequently, in the \textit{AoSA phase}, we examine the involvement of automatic accounts in delivering IoCs. To identify whether an account is an automatic account or a human, we implement ML and DL models. To develop the bot prediction model, we derive diverse features, including profile, temporal, and content levels, from Twitter accounts. Section~\ref{bot_classification} outlines how the features were generated for account classification and the implementation of the ML and DL models. More importantly, we generate interpretable models that explain the features learned by classification algorithms.
\section{Collection and Identification of  Security Tweets Carrying IoCs}
\label{sec:security_tweets}
\subsection{Security Tweets}
Initially, an Twitter developer account was created to access Twitter API services, allowing the extraction of user profile information, block lists, user searches, and real-time tweets. The data collection program automatically harvested 2.41 million tweets(English and non-English) employing the \textit{tweepy}\footnote{https://docs.tweepy.org/en/stable/} python module using keyword/hashtag search. We employed a number of hashtags related to cyberspace, including \textit{malware, ransomware, adware, spyware, malspam, crypto-locker, md5, sha256, phishing,  spearphishing, etc.}~\cite{kristiansen2020cti}~\cite{behzadan2018corpus}~\cite{shin2021twiti}. Since our study focused on tweets written in English,  we excluded non-English tweets from the 2.4 million tweets using the \textit{language} attribute information obtained from the API response. After the elimination of tweets in other languages, we were left with 1.39 million English tweets. Furthermore, our tweet corpus includes a lot of retweets because Twitter offers a retweet feature that permits users to repost previously posted tweets. It is preferable to remove retweets from the collected tweets because they include duplicate text as the original post. Thus, we omitted the retweets using the API response data and received 445,049 original tweets. Although some duplicate tweets may exist in the dataset, it is important to eliminate them to ensure accuracy. As the posting time of each tweet is a crucial element in our study, we sorted all tweets in chronological order and eliminated duplicates by utilizing the \textit{drop\_duplicates()} method of the data frame. Specifically, we kept only the first occurrence of each tweet. After the elimination, \textit{430,187} unique tweets remained.
\subsection{Identification of Relevant Tweets}
The straightforward retrieval of tweets using a keyword-based search strategy yields irrelevant tweets like malware tutorials tweets or security blogs. So, we implemented a tweet classifier to identify the relevant tweets and exclude tweets without pertinent IoCs.
\subsubsection{Preprocessing and Classification} 
 Each tweet is processed through NLP procedures to extract essential information for classification. The preprocessing procedure comprises (i) lowercase conversion, (ii) removal of mentions, (iii) normalization of all the IoCs, including URLs, IPs, domain names, CVEs, and hashes, to their corresponding types, such as [url], [ip], [domain], [cve], [hash], to aid in recognizing tweets that contain IoCs, (iv) substitution of all defanged IoCs\footnote{In order to prevent accidental clicks on harmful links, the IoCs are transformed into another format(i.e., defanged). For example, the defanged representation of http://xn—iconiqcaptal-0fb.com/ is hxxp:/ xn—iconiqcaptal-0fb.com/, while the defanged form of IP address 1.1.1.1 is 1[.][1[.]1[.]1} with [defanged] because a tweet containing a defanged IoC value makes an important feature for classification, (v) removal of non-alpha characters except square brackets, which are used in IoC normalization, (vi) removal of single-character tokens, (vii) lemmatization, and (viii) stopword removal. Later, a deep learning classifier such as 1D-Convolutional Neural Network~(CNN) is used to identify relevant tweets. We employ Glove~\cite{pennington2014glove}, a pre-trained embedding method, to produce numerical feature vectors of 300 dimensions.
\subsubsection{Experimentation}
\textit{\textbf{Experimental Setup:}}
We used Ubuntu 22.04 LTS system with an i9 CPU and 32 GB RAM to carry out this experiment. We used the sci-kit learn and the Keras Python tool to build classification models and \textit{matplotlib} package for visualization.
\par
\textit{\textbf{Dataset}:} 
One Ph.D. student, with the support of two professors with 15 years of experience in cyber security, manually annotated a subset of 32,588 tweets out of 430,187. It is worth noticing that the manually annotated sample ensures a reliable and fair representativeness of the whole data collection (i.e., a confidence level of 99\% and a confidence interval of less than 1\%). The annotators labeled them based on the presence of IoCs in the tweet, giving a label of 1 for relevant tweets and 0 for irrelevant ones. This process resulted in a final dataset of labeled tweets with 12,400 relevant and 20,188 irrelevant tweets. To train the model, 80\% of the labeled data was utilized. The remaining samples were divided equally between validation and testing sets.
\par
 \textit{\textbf{Model Creation and Result:}} 
CNN models have demonstrated their effectiveness in various NLP tasks, including sentence classification, document query matching, and semantic parsing. The ability of CNNs to automatically learn feature representations from input data makes them particularly useful for accurately classifying tweets into different categories. Furthermore, CNNs can handle variable-length inputs by utilizing convolutions of varying sizes and pooling operations to reduce input dimensionality~\cite{stojanovski2015emotion}. So, in this study, we created a 1D-CNN model consisting of six one-dimensional convolution layers, five dropout layers, three max pool layers, and a fully connected layer for tweet classification. Using KerasTuner, we fine-tuned the hyperparameters to determine the ideal values for number filters, kernel size, and learning rate. The resulting model employed convolution layers with filter sizes of 50, 100, and 200 and a kernel size of 3. Finally, we compiled the model using the \textit{Adam} optimizer with a learning rate set to 0.0001. 
\par
  We executed the training for 200 epochs and evaluated the performance. Due to the skewed nature of the dataset, we considered the F1-score and Detection Rate~(DR) to evaluate the performance. The F1-score is the harmonic mean of precision and recall and computed by the Equation~\ref{f1-score}. The DR is the percentage of true positive cases correctly identified by the model among all the actual positive cases and determined by using Equation~\ref{detection_rate}. 
   \begin{equation}
        F1-score=\frac{TP}{TP+\frac{1}{2}(FP+FN)}
        \label{f1-score}
    \end{equation}
    where \textit{True Positive~(TP)} indicates how often the classifier correctly predicts a tweet as relevant. \textit{False Positive~(FP)} represents how many times the model misclassified a tweet as relevant while it is an irrelevant tweet. \textit{False Negative~(FN)} indicates how many times the model misclassified  a tweet as irrelevant while it is a relevant tweet.
    \begin{equation}
        Detection Rate =\frac{TP}{TP+FN}
        \label{detection_rate}
    \end{equation}
  The CNN model achieves an F1-score of 98.80\% and a DR of 99.65\%. After the model was developed, we used the CNN model to predict the label (i.e., relevant or irrelevant) for the unlabeled tweets. As the result of CNN prediction, a total of 58,769 tweets are identified as relevant, and the remaining 371,418 tweets are predicted as irrelevant.
\section{Extraction and Analysis of IoCs using TIS}
\label{sec:ioc_extraction}
\subsection{IoC Extraction}
    The IoC extractor automatically retrieves IoCs from the relevant security tweets by using a regular expression. We adopted the \textit{iocextract}\footnote{https://github.com/InQuest/python-iocextract} library to extract IoCs and handle defanged forms as Twitter defanged the suspicious IoCs to protect users from unintentionally visiting a harmful hyperlink. Additionally, Twitter URLs are often shortened; therefore, in order to improve readability and effectiveness, we expanded the abbreviated URLs obtained as the response from the 
 API response. To create a pertinent IoC dataset, certain unnecessary URLs like Tweet URLs, Facebook and Youtube URLs, and similar ones are also purged.
 \par
Furthermore, it is usual to requote someone else's tweet, and the likelihood of the same IoCs appearing more than once is common. So, we arranged the IoCs chronologically and generated unique IoCs by removing duplicate records. As a result of IoC Extraction, we generated a dataset containing \textit{90,991} IoCs from 58,769 tweets with the following fields: \textit{user name, published date, IoC value, IoC type, hashtags, and tweet URL}.  Table~\ref{tab:IoC_Ststistics} shows how often IoCs were reported on Twitter from January 2021 to July 2021. According to these statistics, IoCs are highly reported in the month of April 2021, and URLs contribute the most. 
\begin{table*}[ht!]
\scriptsize
\caption{Month-wise statistics of various IoCs reported in Twitter}
\centering
   \begin{tabular}{lllllll}
   \hline
      \textbf{Month}& \textbf{URL} & \textbf{IP} & \textbf{Domain}& \textbf{Hash} & \textbf{CVE} & \textbf{Total}\\
      \hline
      \textbf{January} & 8986 & 2708 & 852  & 1602 & 260 & \textbf{14408} \\

      \textbf{February} & 10500 & 1639 & 855 & 565 &	270 & \textbf{13829} \\

      \textbf{March} & 10767 & 1324 & 833 & 302 & 286 & \textbf{13512} \\

      \textbf{April} & 10029 & 2332 & 1350 & 661 & 230  & \textbf{14602} \\

      \textbf{May} & 8457 & 2158 & 918 & 727 & 308  & \textbf{12568} \\

      \textbf{June} & 6440 & 1675 & 927 &	282 & 259 & \textbf{9583} \\

      \textbf{July} & 8134 & 2128 & 1541 & 406 &	280 & \textbf{12489} \\
    \hline
      \textbf{Total IoC} & \textbf{63313} & \textbf{13964} &  \textbf{7276} & \textbf{4545} &\textbf{1893} & \textbf{90991} \\
      \hline
\end{tabular}
\label{tab:IoC_Ststistics}
\end{table*}
\subsection{Meta Data Information about IoC from TIS}
\par
The IoCs extracted from the tweets are queried in various engines such as Virustotal, AlienVault, Urlhaus, Malwarebazaar, MISP, and NVD to determine whether Twitter discloses a significant amount of security vulnerabilities, as well as Twitter gives early warnings about cyber threats. We defined a threshold of one in VirusTotal, which indicates that if any antivirus engine report that an IoC is harmful, then we consider the IoC as malicious. In the case of AlienVault, we used information such as pulse, google safe browsing, and antivirus reports to determine if an IoC is harmful. Both UrlHaus and  MalwareBazaar employ database-based approaches that store a substantial amount of malicious indicators in their database. If the IoCs are found in the  UrlHaus and MalwareBazaar database, we regard them as malicious since both of these engines  do not keep the benign IoCs.
Moreover, MalwareBazaar exclusively focuses on hashes and only gives information about them. All forms of IoCs or attributes, including CVE-IDS, are scannable in MISP through PyMISP. The PyMISP package for Python enables fetching events, adding or modifying events/attributes, and searching for attributes. We can interpret the IoC as a malicious indicator if the attributes are reported in any of the events. Each IoC was submitted to various TISs and a report was prepared that segregates them as clean or malicious. As we previously noted, the NVD and the CVE list are strongly linked, and any updates to the CVE list are immediately replicated in the NVD. However, there is still a time gap between the publishing date in these databases since  CVE is allocated an identifier immediately when a threat occurs, without waiting for additional analysis. We use diverse TIS to collect information in the form of the first reported date, attack category, file types for hashes, etc. These details are also recorded for malicious IoCs in order to compute metrics such as \textit{correctness, timeliness}, and \textit{overlap}. The details about the number of malicious and  clean IoCs obtained from TISs are summarized in Table~\ref{tab:IoC_Status}. 
\begin{table*}[h!]
\scriptsize
\centering
    \caption{The IoC status in various Threat Intelligence Services}
    \scriptsize
    \resizebox{\textwidth}{!}{
   \begin{tabular}{lllllllllllll}
   \hline
   \multirow{2}{*}{\textbf{IoC Type}} &
   \multicolumn{2}{l}{\textbf{VT}} &
   \multicolumn{2}{l}{\textbf{AV}} &
    \multicolumn{2}{l}{\textbf{
    UH}}  & \multicolumn{2}{l}{\textbf{MB}} &
    \multicolumn{2}{l}{\textbf{MISP}} & \multicolumn{2}{l}{\textbf{NVD}} \\
    \hhline{~------------}
      & M & C & M &	C & M & NF & M & NF & M & NF & F & NF \\
      \hline
      \textbf{URL} & 29730 & 33583  & 34 & 63278 & 77 & 63236  & \textbf{NA}& \textbf{NA} & 49 & 63264 & \textbf{NA}& \textbf{NA} \\

      \textbf{IP} & 3035 & 10929  & 1596 & 12368 & 290 & 13674  & \textbf{NA}& \textbf{NA} & 516 & 13448 & \textbf{NA}& \textbf{NA} \\

      \textbf{Domain} & 4108 & 3168  & 183 & 7093 & 167 & 7109  & \textbf{NA}& \textbf{NA} & 63 & 7213 & \textbf{NA}& \textbf{NA}\\

      \textbf{Hashes} & 4417 & 128  & 4223 & 322 & 348 & 4197  & 639 & 3906 & 2004 & 2541 & \textbf{NA}& \textbf{NA}\\

      \textbf{CVE} & \textbf{NA}& \textbf{NA}  & \textbf{NA}& \textbf{NA} & \textbf{NA}& \textbf{NA}  & \textbf{NA}& \textbf{NA} & 101 & 1792 & 1869 & 24 	\\

      \textbf{(\%)} & 46.34 & 53.66  & 6.77 & 93.23 & .99 & 99.01  & 14.06 & 85.94 & 3.00 & 97.00 & 98.73 & 1.27  \\
      \hline
    \end{tabular}}
  \label{tab:IoC_Status}
  $ ^* $ VT- VirusTotal, AV - AlienVault, UH - UrlHaus, MB - MalwareBazaar
\\
     $      ^* $ M- Malicious, C - Clean, F - Found, NF - Not Found, NA - Not Applicable
\end{table*}
Based on our analysis, VirusTotal reports a higher number of IoCs if compared with the other intelligence services. In comparison to others, this platform gives 46.34\% of indicators as malicious and reports almost all file hashes as malicious. When the entire inputs are examined in the AlienVault, it returns only 6.77\% of IoCs as malicious and remaining as clean. Similar to the AlienVault platform, Urlhaus and MISP report a few of the indicators as malicious. However, AlienVault and MISP recognize 92.91\% and 44.09\% of file hashes as malicious, respectively.
MalwareBazar reports that 14.06\% of file hashes retrieved from Twitter are malicious.  Furthermore,  98.73\% of the vulnerability indicators stated on Twitter are already available in the National Vulnerability Database. However, in our analysis, we found 24 CVE IDs (approximately 1.27\% of the total) mentioned on Twitter that are not (yet) included in NVD. We manually verified those 24 CVE IDs in the MITRE CVE list\footnote{https://cve.mitre.org/cve/search\_cve\_list.html} and found that 20 were reserved by an organization or individual for use in announcing a new security vulnerability; one was considered an adversarial sample, and the remaining three were determined to be novel vulnerabilities. These novel vulnerabilities included a credential exposure flaw in the Ansible module~(CVE-2021-20180), and a mutation XSS vulnerability affecting certain combinations of tags with the bleach.clean function in Python~(CVE-2021-23980), and an improper certificate validation vulnerability in Dell EMC Unisphere for PowerMax versions~(CVE-2021-21548).

\section{Generation of Dataset for \textit{AoSA}}
\label{sec:feature_bot}
\label{botdataset}
    \subsection{Feature Generation}
    Compared to generic social media users, accounts delivering threat intelligence information may exhibit odd features (e.g., regularities in the content of tweets, including a high presence of malicious URLs, abnormal social networks, and unusual posting behaviors). In particular, accounts sharing threat intelligence information, such as security organizations, may often follow a predefined schedule or share updates as soon as new information becomes available. Such odd features may lead general-purpose automated solutions to biased bot-detection outcomes.  For instance,  Botometer\footnote{https://botometer.osome.iu.edu/} is a tool which has been used in many studies to detect accounts as human or bot~\cite{alothali2022bot},\cite{akhtar2022machine}. Another tool, Birdspotter, works better than Botometer in terms of mean F1 scores and botness scores~\cite{RamKR21}. It produces high botness value for bot accounts, but some human security accounts such as MBThreatIntel, IronNetTR, s3xcur1ty, CVEannounce, and vigilance\_en have high botness scores (more than 0.99). For this reason, we decided to design a new AI-enabled approach specialized in distinguishing between human-operated and bot-driven accounts posting IoCs. To ascertain this, we derived attributes that are extracted from user profiles and tweet content. The details about the features for the classification of accounts as human or automatic accounts are discussed below.\\ \\
      
 \textbf{\textit{Profile Level Features:}} Profile level features are related to the user's profile information and the account activity. The complete list is provided in Table~\ref{tab:featue_list} and described below.\\
    \textbf{$P_1$: Followers Count: }On social media, a user's number of followers is typically seen as a measure of trust, with genuine followers usually having real-life connections and higher follower counts, while famous personalities differ in this regard, and this can be used to identify potential spammers~\cite{fazil2018hybrid}~\cite{bindu2018discovering}. \\
    \textbf{$P_2$: Following Count: }The authors of~\cite{alom2020deep} utilized the following or friends count~($nF$) as a feature, with spammers commonly following many profiles rapidly to boost their visibility.\\
    \textbf{$P_3$: Followers Ratio: }It represents the number of followers divided by the number of followings, and is indicative of user behavior on social media. Spammers, as observed in~\cite{mateen2017hybrid}, often have fewer followers compared to real users, as they lack real-life connections. Moreover, they tend to follow a larger number of accounts, resulting in a lower ratio, which hints at the presence of bot accounts.
        \begin{equation}
            rFF= \frac{nFoS}{nF}
        \end{equation}
    where nFoS represents the number of
followers of a particular user and nF indicates the number of followings.\\
   \textbf{$P_4$: Listed Count: }The listed count~($nList$) gives the popularity of an Twitter account by estimating the number of users included in other people's lists.
The value of this feature is often high for legitimate users and low for spammers.\\
        \textbf{$P_{5,6}$: Protected and Verified: }Protected~($Pro$) specifies if the profile is private, and verified~($Ver$) shows whether Twitter has verified the account. Spammers often leave their profiles unprotected, and both features have values, either True or False. \\
        \textbf{$P_7$: Length of Description: }This feature, denoted as $lDesc$, measures the character count of the description displayed on the profile page. Automatic accounts, as noted in~\cite{ercsahin2017twitter}, often have blank descriptions, resulting in a description length close to zero.\\
        \textbf{$P_{8}$: Age of Account in Days: }As bot accounts are terminated and disabled by Twitter, the lifespan of spam accounts on Twitter is generally shorter than that of actual users. As a result, spam accounts are typically created more recently than real users~\cite{bindu2018discovering}. The age of the account is the difference between the current date and the profile's creation date on Twitter. 
        \begin{equation}
            age= D_{current}- D_{creation}
        \end{equation}
        \textbf{$P_{9}$: Profile Image: }It indicates the presence of profile image. A bot account is less likely to alter its default profile picture~\cite{ercsahin2017twitter}.\\
        \textbf{$P_{10}$: Listed Count per Day: }It shows how many users add a particular user to their list in a single day, which is expressed using Equation~\ref{eq:listage}:
        \begin{equation}
            rListAge= \frac{nList}{age}
            \label{eq:listage}
        \end{equation}
       \textbf{$P_{11}$: Reputation: }The legitimacy of an Twitter account, whether it is genuine or a bot, can be assessed by calculating its reputation score as described in Equation~\ref{eq:rep}, as noted in~\cite{singh2018twitter}.
        \begin{equation}
            Rep=\frac{nFoS}{nFoS + nF}
            \label{eq:rep}
        \end{equation}
        where $Rep$ indicates the reputation score of the user, $nFoS$ is the number of followers, and $nF$ represents the number of followings. 
\\ \\
\begin{table*}[!ht]
        \centering
        \begin{tabular}{llp{8.5cm}l}
        \hline
           Level & Symbol  &  Feature & Reference \\
           \hline
           Profile & $nFoS$ & Following Count & \cite{fazil2018hybrid},\cite{bindu2018discovering},\cite{alom2020deep},\cite{ilias2021detecting} \\
           Profile & $nF$ & Followers Count & \cite{alom2020deep},\cite{alom2020deep},\cite{ilias2021detecting}\\
           Profile & $rFF$ & Followers Ratio & \cite{mateen2017hybrid},\cite{alom2020deep},\cite{wu2021novel},\cite{ilias2021detecting}\\
           Profile & $nList$ & Listed Count & \cite{ilias2021detecting}\\
           Profile & $lDesc$ & Length of Description & \cite{ercsahin2017twitter},\cite{ilias2021detecting}\\
           Profile & $age$ & Age of Account in Days &\cite{bindu2018discovering},\cite{alom2020deep},\cite{ilias2021detecting} \\
           Profile & $proImage$ & Profile Image & \cite{ercsahin2017twitter}\\
           Profile & $rListAge$ &Listed Count per Day &\cite{ilias2021detecting}\\
           Profile & $Rep$ & Reputation &\cite{singh2018twitter},\cite{alom2020deep},\cite{ilias2021detecting}\\
           Profile & $Pro$ & Protected & \\
           Profile & $Ver$ & Verified &\\
           \hline
           Content & $nTweet$ & Number of Tweet &\cite{fazil2018hybrid}\\
           Content & $nRT$ & Number of Retweet &\cite{fazil2018hybrid}\\
           Content & $\mu_{lenTweet}$& Average length of Tweets&\cite{ilias2021detecting}\\
           Content & $\sigma_{lenTweet}$ & Standard Deviation of Length of Tweets &\cite{rovito2022evolutionary},\cite{ilias2021detecting}\\
           Content & $\mu_{nPurl}$& Average Number of Post Containing URLs &\cite{yang2013empirical}\\
           Content & $nWord$ & Number of Words Per Tweet &\cite{madisetty2018neural},\cite{alom2020deep}\\
           Content & $nDigitTweet$ & Number of Digits Per Tweet &\cite{chen20156},\cite{alom2020deep}\\
           Content & $rMn$ & Username Mention Ratio &\cite{mccord2011spam},\cite{mateen2017hybrid},\cite{ilias2021detecting}\\
           Content & $rUMn$ & Unique Mention Ratio &\cite{amleshwaram2013cats},\cite{ilias2021detecting}\\
           Content & $rHashPost$ & Number of Hashtag per tweet &\cite{ilias2021detecting}\\
           Content & $rURLPost$ & URL ratio &\cite{fazil2018hybrid},\cite{ilias2021detecting}\\
           Content & $rRTPost$ & Retweet Ratio &\cite{wu2021novel},\cite{ilias2021detecting}\\
           Content & $rUPost$ & Unique Post Ratio & \cite{ilias2021detecting}\\
           Content & $\mu_{\alpha}$  & Mean of Like, Quote, Reply, Retweet Count &\cite{kabakus2017survey},\cite{wu2021novel}\\
           Content & $nSoc$ & No of Unique Source &\cite{amleshwaram2013cats}\\
           Content & $rSoc$ & Unique Source Ratio &\cite{wu2021novel},\cite{ilias2021detecting}\\
           Content & ${similarity}$ & Tweet Similarity &\cite{chen2018semi},\cite{ilias2021detecting} \\
           \hline
           Time & $nPostperDay$ & No.of Post Per Day &\\
           Time & $max_{time}$  & Maximum Time Interval Between Tweets, Retweets, Post &\cite{wu2021novel},\cite{ilias2021detecting}\\
           Time & $min_{time}$  & Minimum Time Interval Between Tweets, Retweets, Post &\cite{wu2021novel},\cite{ilias2021detecting}\\
           Time & $\mu_{time}$& Mean of Time Interval Between Tweets, Retweets, Post &\cite{chen2018semi},\cite{wu2021novel},\cite{ilias2021detecting}\\
           Time & $\sigma_{time}$ & Standard Deviation of Time Interval Between Tweets, Retweets, Post &\cite{wu2021novel},\cite{ilias2021detecting}\\
           Time & ${IH}$ & Idle Hours &\cite{alom2020deep},\cite{ilias2021detecting}\\
           Time & ${timePattern}$  & Post Time Pattern &\cite{wu2021novel}\\
           Time & ${burstiness}$ & Burstiness &\cite{pan2016discriminating},\cite{wu2021novel}\\
           \hline
        \end{tabular}
        \caption{List of Extracted features}
        \label{tab:featue_list}
    \end{table*}
\textbf{\textit{Content Level Features:} }%      Social bots typically send similar-sounding tweets with inconsistent grammar and illogical writing patterns. 
We determined the following content-based features to classify Twitter users as human or automatic accounts.\\
        \textbf{$C_{1,2}$: Number of Tweets, Retweets: }Bots often struggle to mimic human tweeting behavior, as they typically generate tweets from databases, use probabilistic methods like the Markov chain algorithm, or opt for retweeting, sending a high volume of tweets to achieve their goals before detection, making this a valuable indicator for identifying spammers. Additionally, the preference of automated spammers for retweets is a distinguishing characteristic, and user behavior is measured in terms of the total number of tweets from the account's creation date~($nTweet$) and the count of retweets~($nRT$) based on tweets starting with RT~\cite{fazil2018hybrid}.\\
        \textbf{$C_{3}$: Average Length of Tweets: }With the use of this feature, we can determine how well a user posts tweets, i.e., around the same length or dynamic length. It computes the average length of tweets posted by a user.
        \begin{equation}
            \mu_{lenTweet}=\frac{\sum_{i=1}^n len(tweet_i)}{N}
        \end{equation}
        where $\mu_{lenTweet}$ is the average length of $N$ tweets, and $len(tweet_i)$ represents length of $i^{th}$ tweet.\\
        \textbf{$C_{4}$: Standard Deviation of Length of Tweets: } In~\cite{rovito2022evolutionary}\cite{CardaioliCSFLV21}, researchers found that automatic accounts have a lower standard deviation of tweet lengths, compared to humans, which implies that the bot accounts are probably posting identical or similar information. This attribute is determined using Equation~\ref{eq:sd}.
        \begin{equation}
            \sigma_{lenTweet}=  \sqrt{\frac{1}{N} \sum_{i=1}^n (tweet_{i}- \mu_{lenTweet})^2}
        \label{eq:sd}
        \end{equation}
       \textbf{$C_{5}$: Average Number of Post Containing URLs: }Spammers generally employ URLs in their tweets, such as advertisements, phishing sites, and redirect links to other websites. The number of tweets, consisting of URLs, is higher among spammers than among legitimate accounts, and this tendency is used to analyze the bot behaviour~\cite{yang2013empirical}\cite{CardaioliCSFLV21}. It is determined by using Equation \ref{avgposturl}.
        \begin{equation}
             \mu_{nPurl}=\frac{n_{Purl}}{nTweet+nRT}
             \label{avgposturl}
        \end{equation}
        where $\mu_{nPurl}$ indicates the average number of posts containing urls, $n_{Purl}$ is the total number posts that contains urls, $nTweet$ indicates total number of tweets, and $nRT$ represents the total number of retweets.\\
       \textbf{$C_{6}$: Number of Words in Tweet: }It is the number of words contained in the tweets for a specific account, denoted by $nWord$. In~\cite{madisetty2018neural}, the researchers suggest that non-spammers employ a significant amount of words in their tweets.\\
        \textbf{$C_{7}$: Number of Digits Per Tweet: }This attribute determines the number of digits per tweet. The bots often use less number of digits compared to  benign users~\cite{chen20156}. 
        \begin{equation}
            nDigitTweet=\frac{n_{digit}}{nTweet}
        \end{equation}
        where $nDigitTweet$ represents the number of Digits Per Tweet, $n_{digit}$ is the number of digits used in the tweets posted by a user, and $nTweet$ is the total number of tweets for the same account.\\
         \textbf{$C_{8}$: Username Mention Ratio:  }Spammers take advantage of the username mentions functionality to propagate spam messages~\cite{mccord2011spam} quickly. A user is regarded as a spam account if their tweets have a lot of mentions and reply tags~\cite{mateen2017hybrid}. The number of mentioned usernames is determined by the API response, represented as $nMnU$. The mention ratio is mathematically represented as: 
        \begin{equation}
            rMn=\frac{nMnU}{nTweet}
        \end{equation}
        where $rMn$ is the mention ratio, $nMnU$ total number of mentioned username, and $nTweet$ represents the total number of tweets.\\
      \textbf{$C_{9}$: Unique Mention Ratio:  }
        Prior studies reported that benign accounts typically post tweets by mentioning a small number of unique usernames~\cite{amleshwaram2013cats}. A high unique mention ratio suggests that the user is engaged in frequent mentions of unique usernames, which signifies that the account is probably a bot. The unique mention ratio is determined by:
        \begin{equation}
            rUMn=\frac{nUMn}{nTweet}
        \end{equation}
        where $rUMn$ represents the unique mention ratio, $nUMn$ total number of unique mentions, and $nTweet$ indicates the total number of tweets.
      \\
       \textbf{$C_{10}$: Number of Hashtag per tweet: }In order to enhance the likelihood that their tweets will be searched, bots frequently include hashtags in their tweets. Therefore, figuring out how many hashtags were mentioned in their tweets adds more significance to the classification phase. It is the ratio of the number of hashtags included in tweets to the number of posts and denoted by $rHashPost$. \\ 
        \textbf{$C_{11}$: URL ratio: }To spread more stuff on Twitter, an automatic account often transmits more URLs in their tweets that sometimes direct to harmful sites. For spammers, inserting URLs is absolutely essential because they cannot fulfill their goals if they do not include URLs in their tweets~\cite{fazil2018hybrid}\cite{CardaioliCSFLV21}. The URL ratio~($rURLPost$), which is the ratio of the number of URLs to the number of tweets, has a high value for spammers but a low value for benign users.\\
      \textbf{$C_{12}$: Retweet Ratio: }The retweet ratio~($rRTPost$) is the proportion of the number of retweets to the total number of tweets and retweets posted by a specific user. 
        \begin{equation}
        rRTPost= \frac{nRT}{nTweet + nRT}
        \end{equation}
        where $rRTPost$ represents the retweet ratio, $nRT$ indicates the number of retweets for a specific user, and $nTweet$ \& $nRT$ are the number of tweets and retweets posted by a user.\\
        \textbf{$C_{13}$: Unique Post Ratio: }Spammers frequently retweet and send similar tweets. As a result, a spammer has fewer unique tweets than a benign tweeter. 
        \begin{equation}
            rUPost=\frac{nUPost}{nTweet+nRT}
        \end{equation}
        where $rUPost$ represents the Unique Post Ratio, $nUPost$ is the total number of unique posts, and $nTweet$ and $nRT$ indicates the total number of tweets and retweets.\\
        \textbf{$C_{14,15,16,17}$: Mean of Like, Quote, Reply, Retweet Count: } It is possible that the spammers' tweets will receive fewer retweets and likes than those of genuine users because their tweets are unsolicited in nature. Furthermore, as real users often avoid tweets from bot accounts, posts of an automatic account receive fewer replies, quotes, and mentions~\cite{kabakus2017survey}.
        \begin{equation}
            \mu_{\alpha}= \frac{\sum_{i=1}^n \alpha_{i}}{n}
        \end{equation}
        where $\mu_{\alpha}$ denotes the mean of Like/ Quote/ Reply/ Retweet Count, $n$ indicates the total number of tweets of a particular user, $\alpha_{i}$ represents the number of Like/ Quote/ Reply/ Retweet obtained for the $i^{th}$ tweet.\\
        \textbf{$C_{18}$: No of Unique Source: }The tweets can be posted through various sources like an API interface, a web application, an HTTP interface, a blogging platform, etc. A legitimate user might not always limit publishing tweets  from a specific source. But the bots could limit their tweets to a few sources because of scale and automation-related considerations~\cite{amleshwaram2013cats}. So, the number of unique sources~($nSoc$) used by accounts gives insight into the botness level. The more number of unique sources implies the more likely the account is to be genuine.\\
        \textbf{$C_{19}$: Unique Source Ratio: }It is the ratio between the total number of unique sources used by one user and the total number of unique sources used by all users. Since the benign users use multiple sources, the ratio will be higher for the real users. It is defined as:
        \begin{equation}
            rSoc=\frac{\text{No. of Unique Source of user}}{\text{No. of Unique Source of all users}}
        \end{equation} 
        \textbf{$C_{20}$: Tweet Similarity: }The tweet's content determines the account's spamming behavior. If an account's tweets are similar, it is more likely that it is a spammer~\cite{chen2018semi}. The cosine similarity of the tweets measures how closely the tweets are related, and a greater score for this feature suggests that the account could be a spam account.
        \begin{equation}
            similarity= \frac{tweet_{i} \cdot tweet_{j}}{|| tweet_{i} || * || tweet_{j} ||}
        \end{equation}
        where $tweet_{i} \cdot tweet_{j}$ indicates dot product of two tweets, and $|| tweet_{i} ||$ indicates norm.\\ \\
   \textbf{\textit{Time Level Features:} }
    The temporal distribution of postings has been seen to differ among spammers and legitimate users. We extracted 16 temporal features.\\ \\
   \textbf{$T_{1}$: No.of Post Per Day: }The amount of tweets and retweets shared by spammers in a day is often higher than that of non-spammers. Therefore, we adopted this feature for identifying bot users, and it is the ratio of the number of posts~(both tweets and retweets) to their account age. 
        \begin{equation}
            nPostperDay=\frac{nTweet+nRT}{age}
        \end{equation} 
    \textbf{$T_{2,3.4}$: Maximum Time Interval Between Tweets, Retweets, Post: }Spammers typically may not tweet for a prolonged amount of time after publishing a significant number of posts in a short period of time~\cite{wu2021novel}. So, we considered the minimum and the maximum time interval between tweets as a feature to classify bots from non-bot users.
    \begin{equation}
        max_{time}=max(T_i-T_j)
    \end{equation}
    where $max_{time}$ is the maximum time interval between two consecutive tweets, retweets, or both tweets and retweets.\\
     \textbf{$T_{5,6,7}$: Minimum Time Interval Between Tweets, Retweets, Post: }Similar to the maximum time interval between tweets, the minimum time gap between consecutive tweets is a significant factor for bot classification~\cite{wu2021novel}. 
    \begin{equation}
        min_{time}=min(T_i-T_j)
    \end{equation}
    where $min_{time}$ indicates the minimum time gap between two consecutive tweets, retweets, or both tweets and retweets.\\
     \textbf{$T_{8,9,10}$: Mean of Time Interval of Tweets, Retweets, Post: }Bot users constantly disseminate spam data, whereas human individuals tweet at random times, typically during their leisure time~\cite{chen2018semi}. The average duration between tweets has the potential to affect the behavior of the bot.
    \begin{equation}
        \mu_{time}=\frac{1}{N} \sum_{i=1}^n (T_{i}-T_{j})
    \end{equation}
    \textbf{$T_{11,12,13}$: Standard Deviation of Time Interval Between Tweets, Retweets, Post: }Like mean of the time interval of tweets, Wu et al.~\cite{wu2021novel} also employed this feature in their study. Hence, we also used the standard deviation of the time interval between tweets to accomplish bot detection.
    \begin{equation}
        \sigma_{time}=\sqrt{\frac{1}{N} \sum_{i=1}^n (\Delta_{i}- \mu_{time})^2}
    \end{equation}
    where $\sigma_{time}$ is the standard deviation of time interval between tweets, $\Delta_{i}$ indicates time interval between two consecutive tweets, and $\mu_{time}$ represents the mean of time interval.\\
   \textbf{$T_{14}$: Idle Hours: }The one objective of an automatic account is to post messages continuously. So, determining the idle hours of a user aids in examining the behavior of an automatic account. The idle hours are determined by the following formula, and it is typically lower for the bot accounts~\cite{ilias2021detecting}.
    \begin{equation}
        IH=\frac{max(T_i - T_j)}{nTweet+nRT}
    \end{equation}
    \textbf{$T_{15}$: Post Time Pattern: }The majority of bots were discovered to automatically publish posts(both tweets and retweets) at predetermined intervals using the web interface or Twitter APIs~\cite{wu2021novel}. This trend suggests a reduced entropy for the duration between tweets. The post-time pattern is determined using the Shannon entropy mechanism, and the formula is provided below.
    \begin{equation}
        timePattern=-\sum_{i=1}^k \mathcal{P}(\Delta_{i}) * log(\Delta_i)
    \end{equation}
    where $\mathcal{P}(\Delta_{i})$ is the probability of $i^{th}$ time interval. \\
  \textbf{$T_{16}$: Burstiness: }Human activities show significant diversity, which means that people sometimes tweet quickly and some other times take a longer period between tweets~\cite{pan2016discriminating}. This nature is determined by the burstiness feature, and it is mathematically represented as follows:
\begin{equation}
    burstiness=\frac{\sigma_{time} -\mu_{time}}{\sigma_{time}+\mu_{time}} +\epsilon
\end{equation}
where $\epsilon$ is used to ensure that the burstiness value is non-negative, the value of the burstiness maybe $\epsilon$ or $\epsilon-1$ or $\epsilon+1$. If the value is $\epsilon-1$, it reflects entirely normal behavior, $\epsilon$ denotes totally Poisson activity, and $\epsilon+1$ indicates the most bursty conduct. In general, spammers tend to follow values close to $\epsilon-1$  and $\epsilon+1$. 
  \begin{figure*}[ht]
    \centering % <-- added
\begin{subfigure}{0.45\textwidth}
  \includegraphics[width=\linewidth,height=4.9cm]{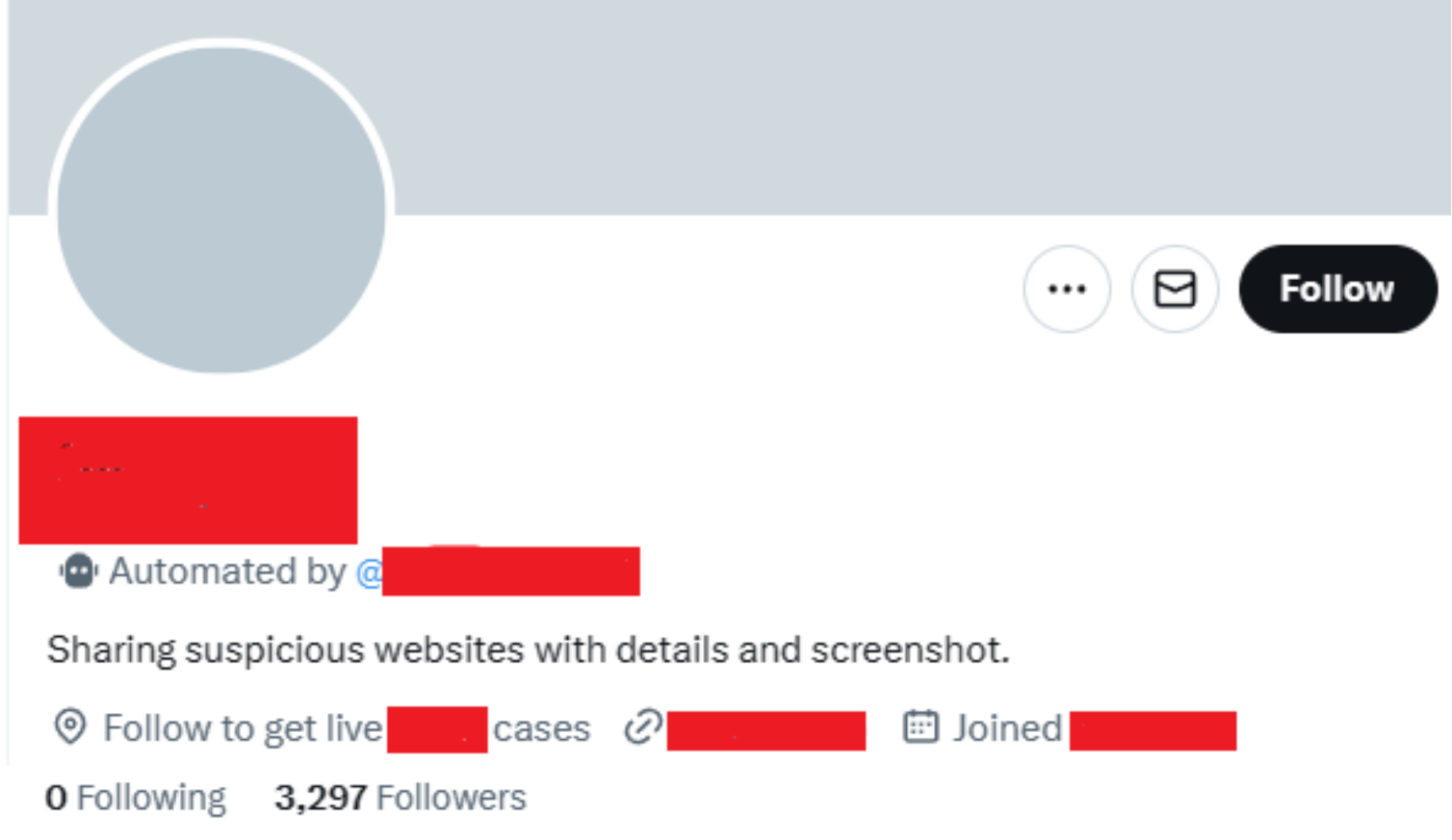}
  \caption{An automated account labeled as \textit{`Automated by'}}
  \label{fig:bot_sample_1}
\end{subfigure}
\begin{subfigure}{0.45\textwidth}
  \includegraphics[width=\linewidth,,height=5cm]{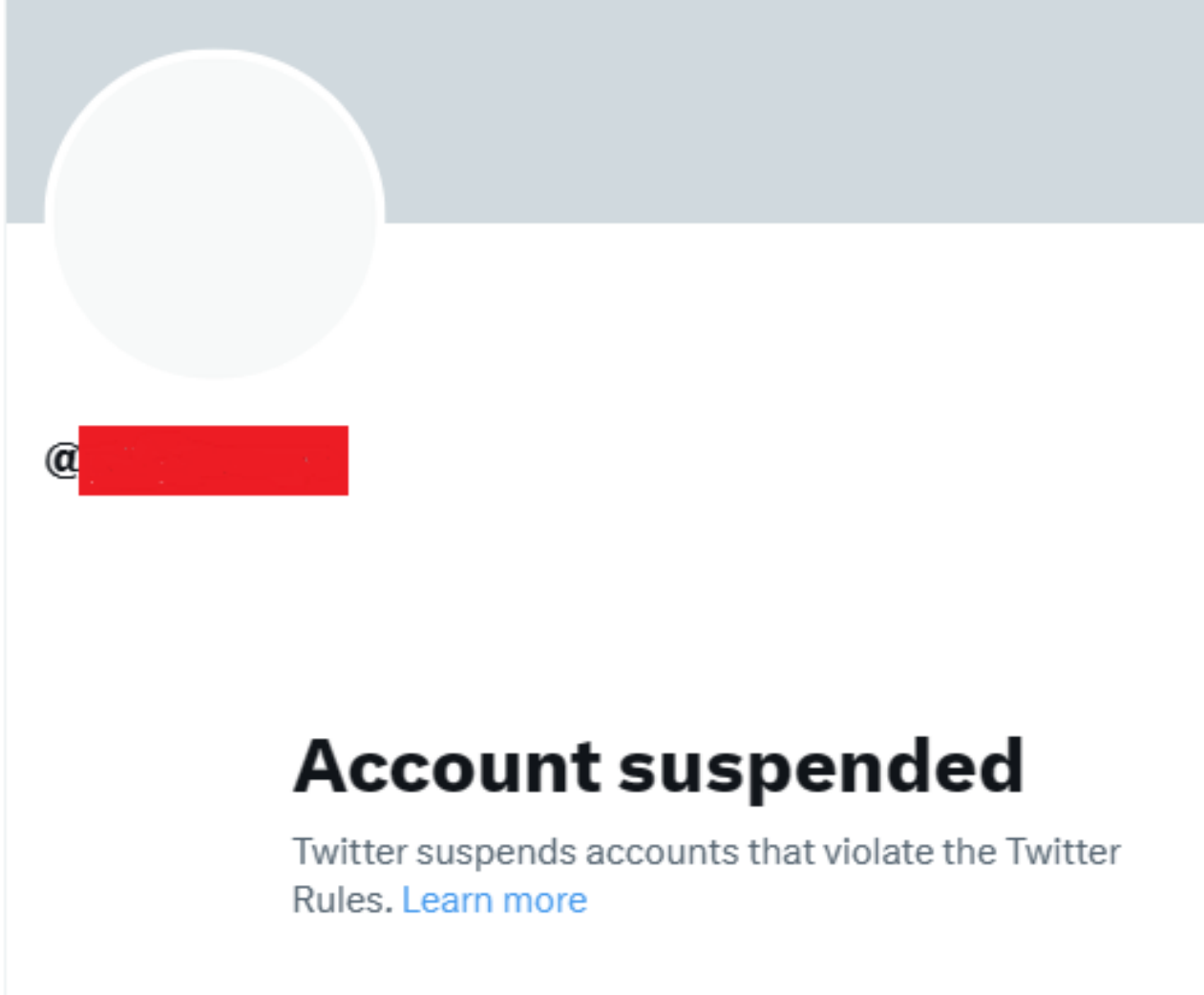}
  \caption{An example of a Suspended Account reported in Twitter}
   \label{fig:bot_sample_2}
\end{subfigure}\hfil % <-- added

\caption{ Profile information of example accounts validated through Twitter}
\label{fig:bot_samples}
\end{figure*}
\subsection{Data Labeling}
  The feature generation phase delivers a data set with 47 features. To build supervised ML/DL models, we performed manual labeling on a subset of the 202,996 security accounts that contained threat information in the given timespan. Due to the labor-intensive and time-consuming nature of manual labeling, which involves reviewing extensive textual content such as tweets and descriptions and considering various non-textual user-related aspects, we randomly selected a statistically significant sample (i.e., a confidence level of 99\% and a margin of error lower than 3) of 3,907 accounts for manual labeling. The labeling was done by a Ph.D. student and two professors with expertise in cybersecurity, all authors of the paper. Initially, accounts were manually labeled as belonging to six categories: \textit{human, company, does not exist (anymore), suspended, bot, and automated by}. We made decisions based on factors such as the profile picture, account description, location, and tweet activities. We were able to clearly identify accounts that belonged to the \textit{human} category by examining their profile pictures and descriptions. However, it was also noted that some accounts might be maintained by companies, and we labeled them as \textit{company} accounts. We discovered that a number of accounts had been suspended by Twitter, did not exist (anymore), or were flagged as ``Automated by”. Figure~\ref{fig:bot_sample_1} shows an instance of an account labeled as ``Automated by”. These accounts were labeled as automated because Twitter allows identifying bot accounts using the keyword ``Automated by” in their profile, as documented in\footnote{https://help.twitter.com/en/using-twitter/automated-account-labels}. We also encountered suspended accounts (like the one depicted in Figure~\ref{fig:bot_sample_2}). Twitter mentions an account may be suspended if it is spammy, hacked, or creates abusive tweets\footnote{https://help.twitter.com/en/managing-your-account/suspended-twitter-accounts}. Additionally, some accounts explicitly stated themselves as bots in their description, and we labeled them as belonging to the \textit{bot} category. For the purpose of bot detection, we considered an account as human-operated if it was associated with a \textit{human} or a \textit{company}.
   Additionally, we view an account as automated if it was \textit{suspended} or identified as a \textit{bot} or \textit{``Automated by"}. Because it is challenging to distinguish between a \textit{does not exist} account as either a human or automated one, we excluded these samples when generating the model. After annotation, we generated a dataset with 47 attributes that included 3,231 legitimate users and 452 automated accounts. 
\begin{figure*}[ht]
       \centering
\includegraphics[width=.8\linewidth,keepaspectratio]{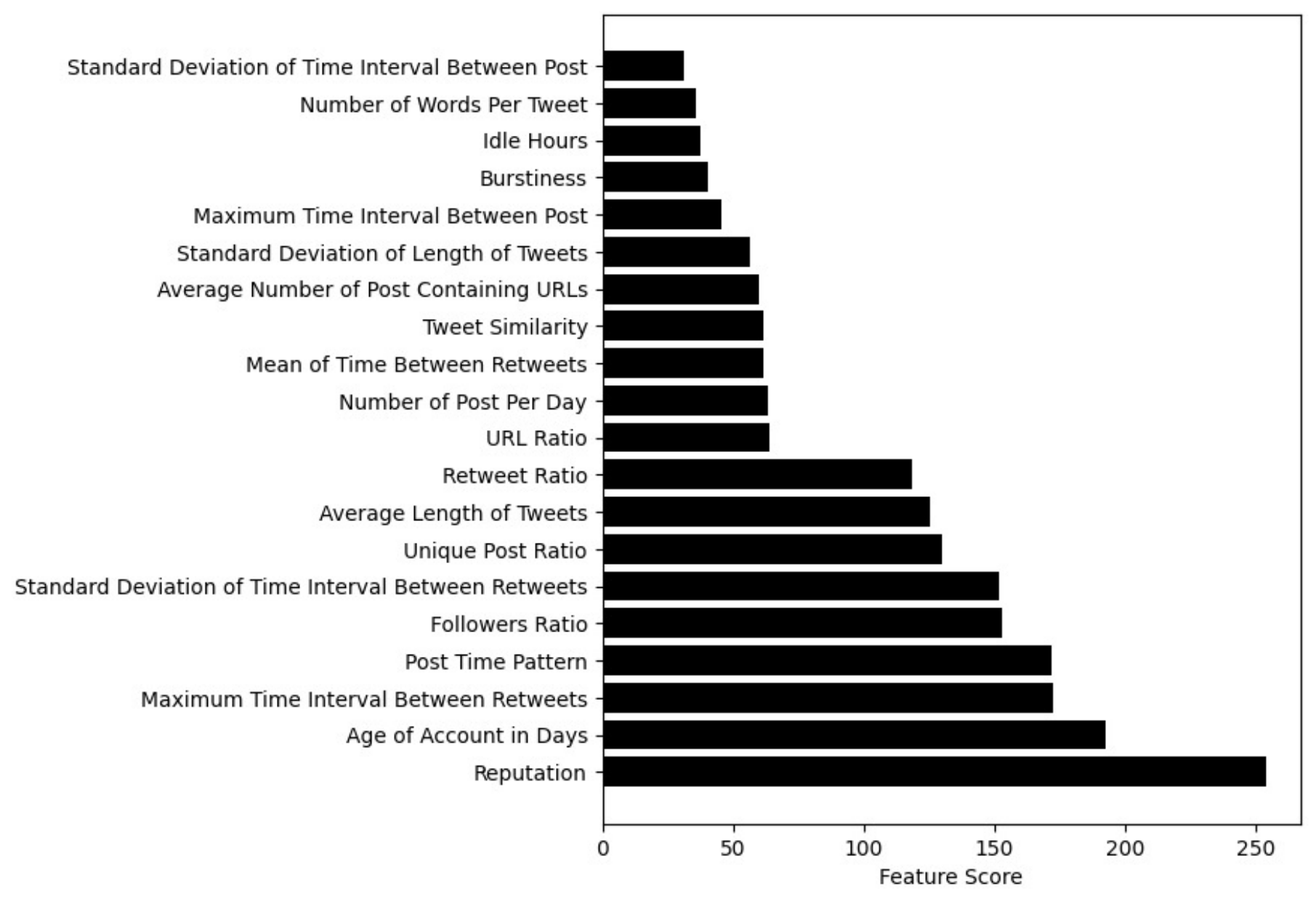}
       \caption{The most relevant 20 features computed using \textit{SelectKBest} method}
       \label{fig:kbestfeature}
   \end{figure*}
\subsection{Data Preprocessing}
Data preprocessing is an essential step in ML models because the quality of the data and relevant information has a direct impact on model performance. So we have performed preprocessing techniques such as data encoding and data normalization. The data normalization helps in transforming the categorical attributes into a machine-understandable format. Our dataset contains two categorical attributes: protected and verified.  So, we converted the verified and protected boolean values to binary by mapping the False to 0 and the True to 1. Furthermore, the min-max normalization is performed on the dataset to scale each feature value between 0 and 1.
\section{Evaluation of AoSA using Classification Models}
\label{bot_classification}
\subsection{Feature Selection}
The model performance is adversely affected by redundant or correlated features; therefore, feature selection techniques aid in filtering out irrelevant features. Feature selection minimizes storage requirements and prevents overfitting by eliminating unnecessary features. We have employed the \textit{SelectKBest} feature selection technique, which outputs the attributes according to the $k$ highest score~($k$ value defines the number of best features). The value of $k$ obtained in one classifier is not appropriate for another because each ML classifier has a unique nature and operates differently depending on various parameters. Therefore, we applied the \textit{SelectKBest} approach to each machine classification model. Additionally, we evaluated the effectiveness of the classifier for various values of $k$(1 to 47). We fix the $k$ for a particular classifier by assessing the performance of the classifier. Conversely, there is no need to perform feature selection on deep learning models since DL classifiers inherently extract pertinent features in their layers.
\subsection{Classification Algorithms}
ML algorithms can learn from a more diverse range of characteristics and data points than heuristic-based methods, resulting in greater precision in identifying accounts driven by bots. The self-build classifier can adapt to add features and changes more easily and identify new patterns of bot behavior. We employed six ML classifiers and one DL model to classify IoC-related Twitter accounts as bot-driven or human-operated. With the help of various ML and DL classifiers such as K-Nearest Neighbor~(KNN), Logistic Regression~(LR), Support Vector Classifier~(SVC), Decision Tree~(DT), Random Forest~(RF), XGBoost, and Deep Neural Network~(DNN), we determined whether Twitter account as human or automatic account.
\subsection{Experimentation Result}
The dataset was split into an 8:2 ratio, with 80\% used for training and 20\% reserved as the test set to evaluate the model's performance. To enhance the models' performance, we fine-tuned the parameters using different methods. Specifically, we employed RandomizedSearchCV for XGBoost and RF, GridSearchCV for DT, KNN, LR, and SVM, and leveraged the KerasTuner mechanism to discover optimal hyperparameters for the DNN.
The selected hyperparameter values are summarized in Table~\ref{tab:hyperparameter}.
\begin{table*}[!ht]
    \centering
     \begin{tabular}{lp{15cm}}
     \hline
         \textbf{Model} &  \textbf{Hyper Parameter}\\
         \hline
         KNN & metric=minkowski, n\_neighbors=5, p=2,  weights=uniform\\

         LR & C=100, penalty=l1, solver=liblinear\\

         SVC & C=400, class\_weight=balanced, gamma=0.01, kernel=rbf\\

         DT & max\_depth=10, max\_features=0.4, min\_samples\_split=10\\
 
         RF & class\_weight=balanced\_subsample, max\_depth=110, min\_samples\_leaf=4, min\_samples\_split=10,                     n\_estimators=2800\\

         XGBoost & {reg\_lambda = 12.8, reg\_alpha =0.2, n\_estimators = 2800, min\_samples\_split = 5, min\_samples\_leaf = 1, max\_features = sqrt, max\_depth = 70, learning\_rate = 0.06, gamma = 0.8} \\

         DNN & optimizer = adam, learning\_rate = 0.001, number of neurons: 47, 32, 8\\
         \hline
    \end{tabular}
    \caption{List of Hyper Parameter Values}
    \label{tab:hyperparameter}
\end{table*}
   We implemented six ML models and one DL model and subsequently assessed their performance using the F1-score as the evaluation metric. We opted for the F1-score over accuracy due to the class imbalance within the dataset, specifically, the presence of fewer automated accounts in comparison to human-operated samples. To enhance classifier performance, we conducted feature selection for each ML classifier, determining the optimal number of features denoted as $k$. Table~\ref{tab:performance} presents the performance results in two scenarios: one with the complete set of features and the other with only $k$ selected features. After feature selection, certain classifiers exhibited a marginal improvement in their F1 scores. Furthermore, we explored how well the number of important features differs across classifiers. For SVM, the best results were achieved when selecting 33 features, resulting in a macro F1 score of 0.763 and a weighted F1 score of 0.887. Similarly, DT, KNN, and LR demonstrated strong performance with 31, 16, and 37 attributes, respectively. In contrast, RF and XGBoost achieved their highest F1 scores when retaining nearly all features, around 43, and 46 respectively. The ranking of features according to the \textit{feature score} obtained from the SelectKBest algorithm is depicted in Fig~\ref{fig:kbestfeature}. 
   The most relevant features found are reputation, account age, the maximum time interval between tweets, post-time pattern, etc. From Table~\ref{tab:performance}, we infer that XGBoost achieved superior performance with a macro F1-score of 0.814 and a weighted F1-score of 0.925  compared to other classifiers. Using the XGBoost model, we predicted \textit{human/bot} labels for 202,996 security accounts that had contributed to the 2.4 million tweets collected via the hashtag mechanism. Our predictions revealed that within this dataset, a substantial portion of 198,026 accounts are classified as human-operated, whereas a relatively smaller fraction, encompassing 4,970 accounts, have been identified as automated accounts. Section~\ref{exp} explains which features influenced the correct classification and misclassification. 
\begin{table*}[!ht]
    \centering
    \scriptsize
    \caption{Performance of ML and DL models with the full feature set and best relevant features}
    \resizebox{\textwidth}{!}{
    \begin{tabular}{llllll}
    \hline
       \multirow{2}{*}{Model} & \multirow{2}{*}{Macro F1-score} & \multirow{2}{*}{Weighted F1-score} &  \multicolumn{3}{l}{Feature Selection}              \\  \hhline{~~~---} 
                    &                     &  & \multicolumn{1}{l}{k value} & \multicolumn{1}{l}{ Macro  F1 score} & Weighted F1 score \\ \hline
         SVM & 0.754	& 0.882	& 33	& 0.763	 &0.887\\

        DT & 0.768	& 0.907	& 31	& 0.771	& 0.907\\

        KNN & 0.774	& 0.912	& 16	& 0.807	& 0.924\\
    
        LR & 0.787	& 0.914	& 37	& 0.809	& 0.924\\

        RF & 0.81 & 0.922 &	43 & 0.81 &	0.922 \\

           DNN  & 0.812 & 0.922 & NA & 0.812 & 0.922\\

        XGBoost & 0.814	& 0.925	& 46	&\textbf{ 0.814}	& \textbf{0.925}\\
        \hline
    \end{tabular}}
    \label{tab:performance}
\end{table*}
\\
\textit{\textbf{Comparison with Birdspotter}}: We conducted a comparative analysis of our model against the Birdspotter solution. The Birdspotter tool assigns a botness score to each Twitter account, ranging from 0 to 1. We computed botness scores for all Twitter accounts in our test dataset using this tool. A botness score near 1 signifies a strong likelihood of being a bot, whereas a score closer to 0 indicates a lower likelihood. To facilitate our comparison, we established a threshold of 0.95, designating accounts with scores at or above this threshold as bot~(labeled as 1), while those below were considered human~(labeled as 0). Subsequently, we computed the classification performance of the Birdspotter threshold-based solution, using manual labels as the ground truth. The results showed 71 True Positives, 172 False Positives, 19 False Negatives, and 475 True Negatives. Therefore, the Birdspotter-based solution yielded a macro F1 score of 0.63, a weighted F1 score of 0.783, a recall of 0.73 for the human class, and a precision of 0.29 for the bot class. In contrast, the XGBoost model yielded a performance of 52 True Positives, 14 False Positives, 633 True Negatives, and 38 False Negatives. In particular, %looking at the XGBoost results,
compared to the Birdspotter-based approach, the XGBoost model produces fewer false positives, achieving a higher precision on the bot class~(0.79) and higher recall on the human class~(0.96). This finding not only confirms our hypothesis that a general-purpose state-of-the-art bot identification solution (i.e., Birdspotter) can easily confound a human-operated account spreading IoC information with a bot, but also demonstrates that our model's classification outputs are more reliable than the ones obtained through Birdspotter when dealing with CTI-related accounts.
\begin{figure*}[!ht]
    \centering
\includegraphics[width=0.7\linewidth,keepaspectratio]{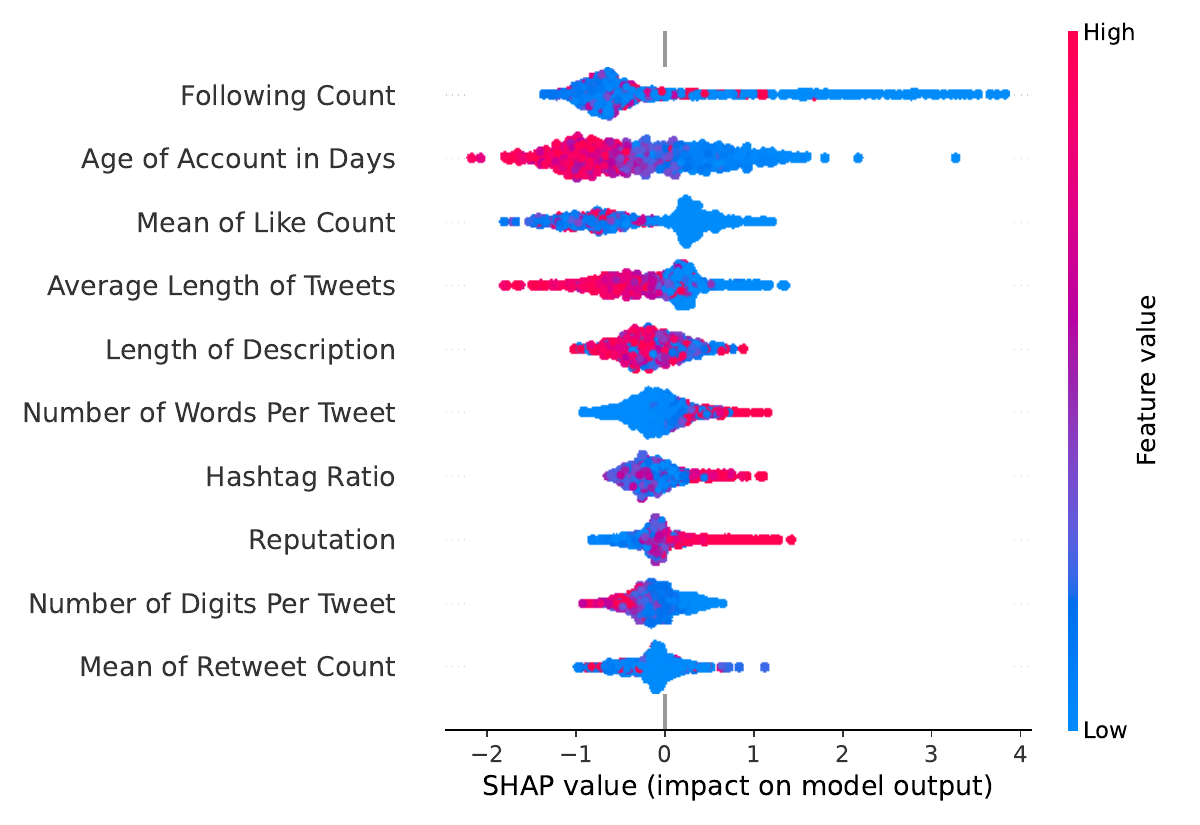}
    \caption{Global Interpretation of Model Using SHAP}
    \label{fig:shap_summary_plot}
\end{figure*}
\begin{figure*}
    \centering % <-- added
  \includegraphics[width=\linewidth]{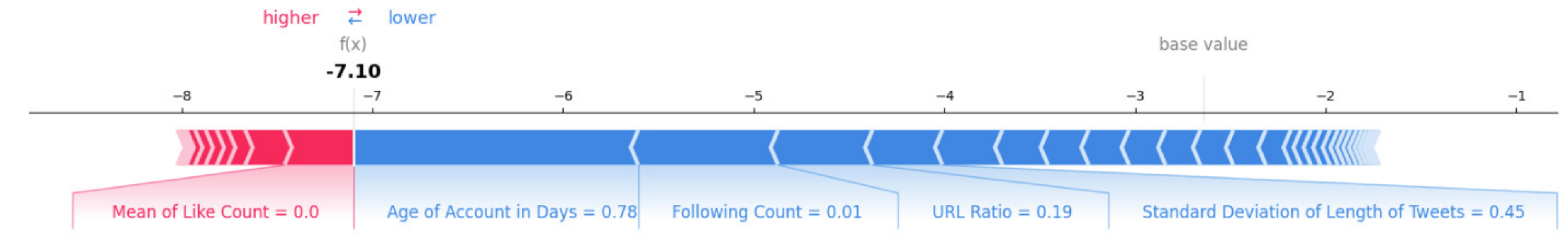}
  \caption{Local Interpretation of correctly classified \textit{legitimate account} using SHAP force plot}
  \label{fig:shap_sample_1}
\end{figure*}
\subsection{Model Explainability using SHAP and LIME}
\label{exp}
An explainable artificial intelligence provides information or justifications to make the model functioning obvious or simple to comprehend~\cite{arrieta2020explainable}. We adopted the SHapley Additive Explanation~(SHAP) and Local Interpretable Model-Agnostic Explanation~(LIME) approaches to explain the bot classification model and determine the features that influence model prediction. SHAP and LIME are two model-agnostic techniques for explaining any classifier. SHAP~\cite{lundberg2017unified} is based on the coalition idea of game theory and provides a means to assess the influence of each attribute. LIME~\cite{ribeiro2016should} concentrates on explaining a particular instance locally. Since the XGBoost model performed well in classifying accounts, we interpreted and explained the XGBoost model through different explainability techniques.
\subsubsection{Global Interpretability}
Global interpretability facilitates grasping the complete logic behind a model and each of the various possible outcomes. In this phase, we exploited SHAP to understand and explain the XGBoost model globally by determining SHAPley values. Figure~\ref{fig:shap_summary_plot}  illustrates the SHAP summary Beeswarm plot, which displays the features according to 
their relevance in classifying Twitter accounts. The horizontal axis indicates the SHAP value, while the vertical axis indicates the features in increasing order of relevance. Each of the tiny points on the figure corresponds to a single instance. We can determine whether an instance has a greater or smaller feature value by looking at the color of the particular dot. The point is red when the feature has high values and blue for those with relatively low values. From Figure~\ref{fig:shap_summary_plot}, we observe that some very low values of the \texttt{following count} have highly impact on the model prediction. Similarly, the higher values of \texttt{account age} negatively correlated with the model's score. But in the case of \texttt{mean of retweet count}, a few low values of this feature have a little detrimental influence on performance.
\begin{figure*}
    \centering % <-- added
  \includegraphics[width=\linewidth]{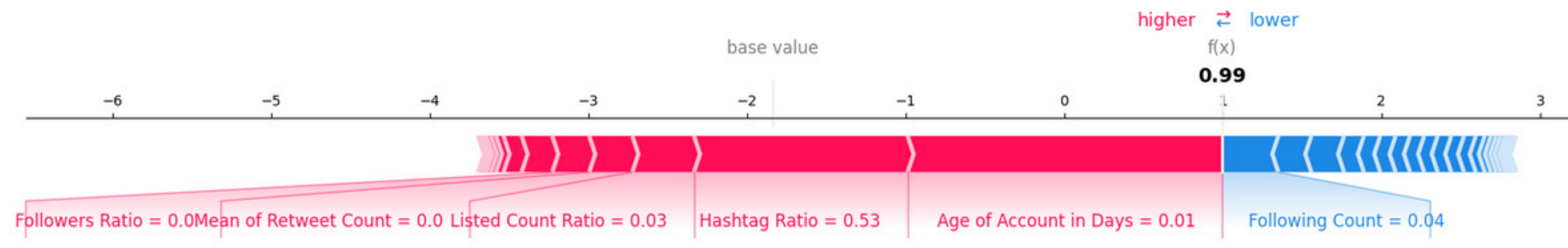}
  \caption{Local Interpretation of correctly classified \textit{bot account} using SHAP force plot}
  \label{fig:shap_sample_2}
\end{figure*}
\subsubsection{Local Interpretability}
\begin{figure*}[ht]
    \centering % <-- added
  \includegraphics[width=\linewidth]{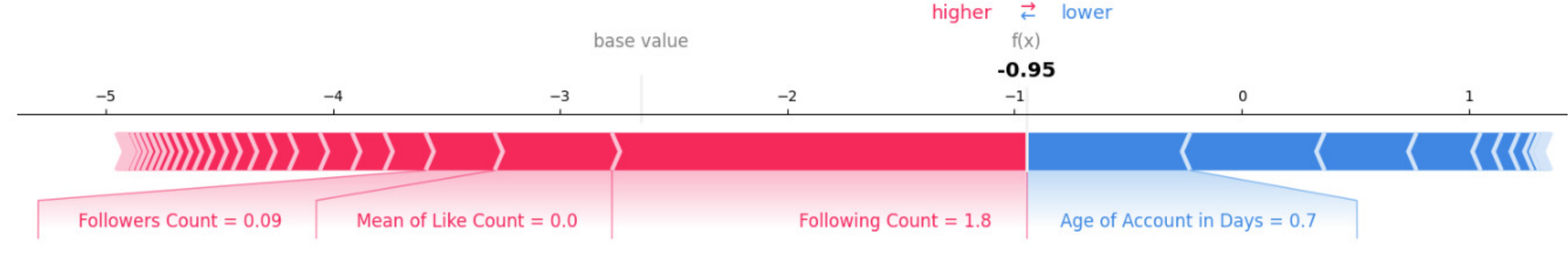}

  \caption{Local Interpretation of misclassified data sample using SHAP force plot}
  \label{fig:shap_sample_3}
\end{figure*}

Local Interpretability aids in comprehending and interpreting a prediction regarding a particular data instance. We applied both SHAP and LIME to interpret the given instance. Figure~\ref{fig:shap_sample_1}~\ref{fig:shap_sample_2}~\ref{fig:shap_sample_3} depicts a SHAP force plot that outlines how each attribute affects the prediction of a given sample. The base value indicates the mean of all predicted probabilities. The prediction for the given data is shown in bold. The hues blue and red signify the key attributes used to generate the prediction for the given instance. Red denotes attributes that increase the model score for the particular data instance, whereas blue represents features that diminish the model score. The most influential features for the prediction are located near the red-blue borderline, and the width of the color bar represents the influence of that feature in the prediction score.
\par
 Figure~\ref{fig:shap_sample_1} demonstrates the SHAP interpretation for a specific legitimate account. We can analyze that the predicted value for this sample is -7.10, which is smaller than the base value, so it is predicted as human. The feature values highlighted in red, in this case, \texttt{mean of like count}, have been shown to boost the model score. Also, features like \texttt{account age, following count, url ratio, and standard deviation of length of tweets} decreases the model's score. The local interpretation of a specific automated account is illustrated in Figure~\ref{fig:shap_sample_2}. The output value mentioned in the plot is 0.99, which is higher than the base value. The features shown in red have a beneficial impact on the output value, with the \texttt{account age} having the highest influence.
\par
\begin{figure*}[ht]
    \centering % <-- added
\begin{subfigure}{0.45\textwidth}
  \includegraphics[width=\linewidth]{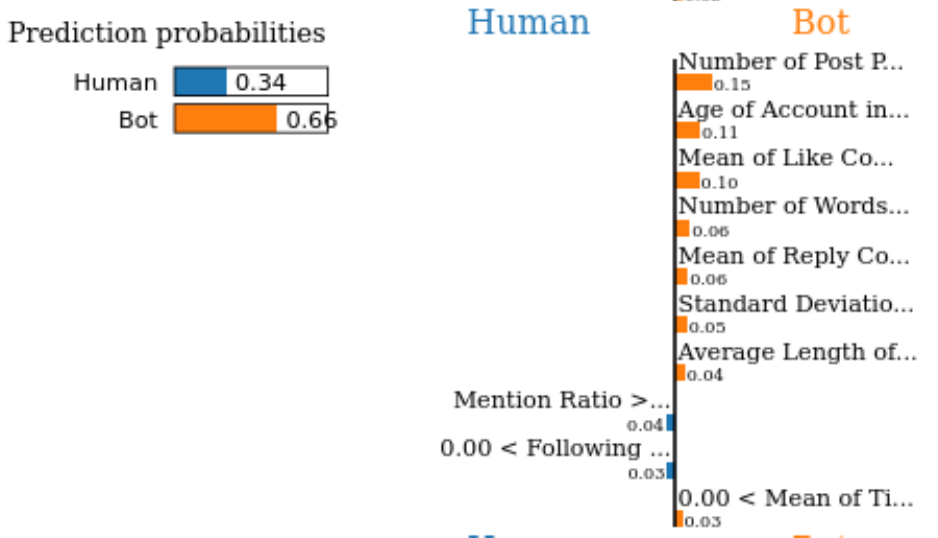}
\end{subfigure}\hfil % <-- added
\begin{subfigure}{0.38\textwidth}
  \includegraphics[width=\linewidth]{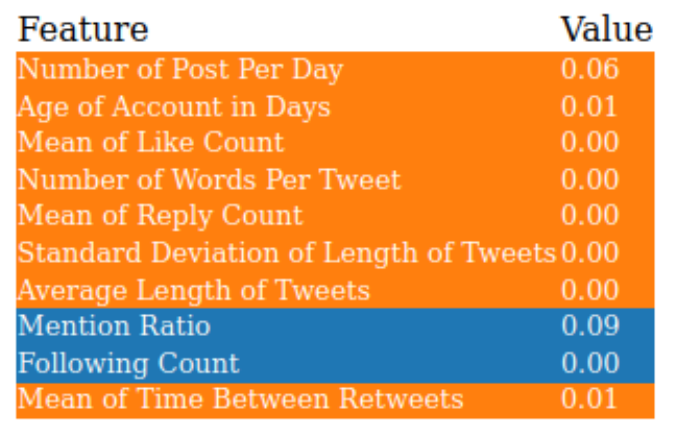}
\end{subfigure}
  \caption{Interpretation of a specific bot account using LIME}
  \label{fig:lime_sample_1}
\end{figure*}
\begin{figure*}[ht]
    \centering % <-- added
\begin{subfigure}{0.45\textwidth}
  \includegraphics[width=\linewidth]{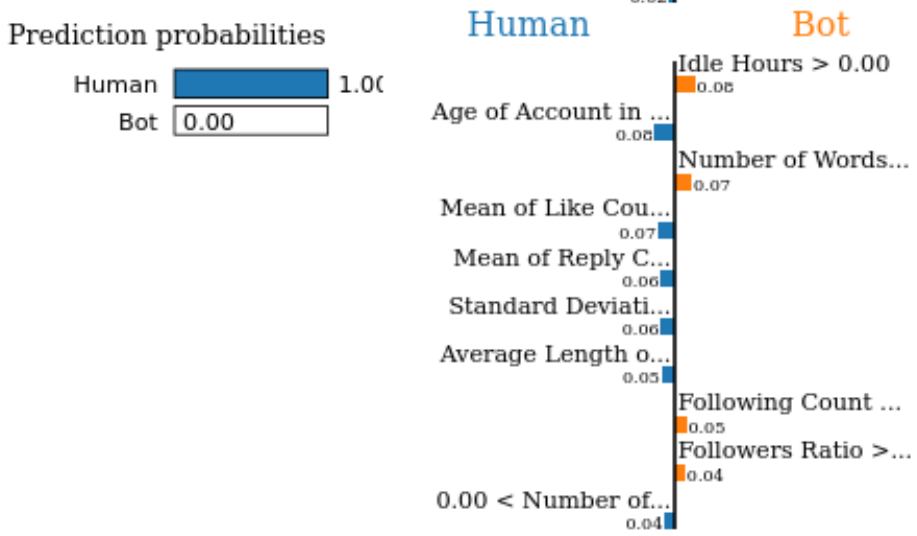}
\end{subfigure}\hfil % <-- added
\begin{subfigure}{0.38\textwidth}
  \includegraphics[width=\linewidth]{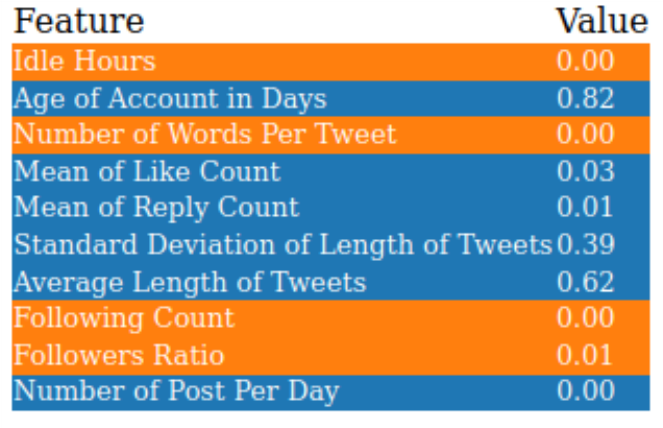}
\end{subfigure}
  \caption{LIME Interpretation of a correctly classified human account}
  \label{fig:lime_sample_2}
\end{figure*}
\begin{figure*}[ht]
    \centering % <-- added
\begin{subfigure}{0.45\textwidth}
  \includegraphics[width=\linewidth]{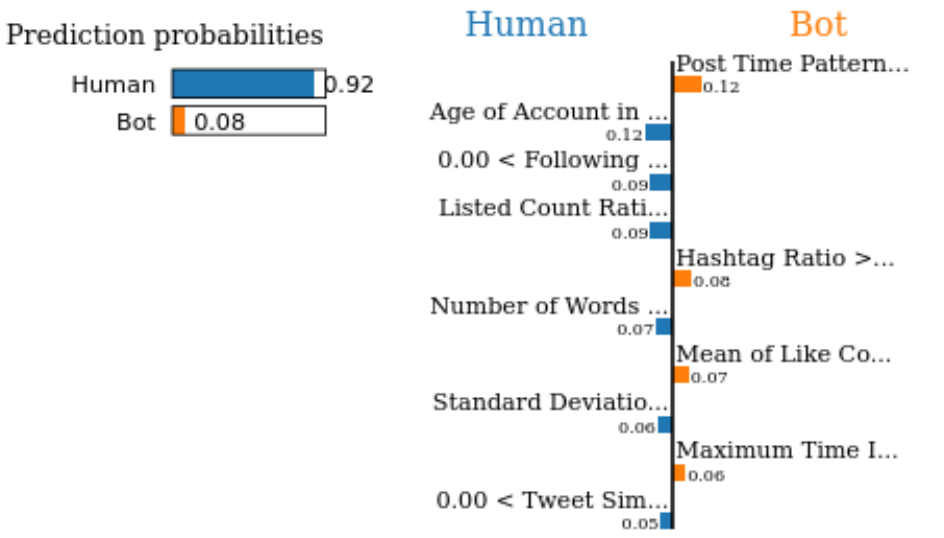}
\end{subfigure}\hfil % <-- added
\begin{subfigure}{0.47\textwidth}
  \includegraphics[width=\linewidth]{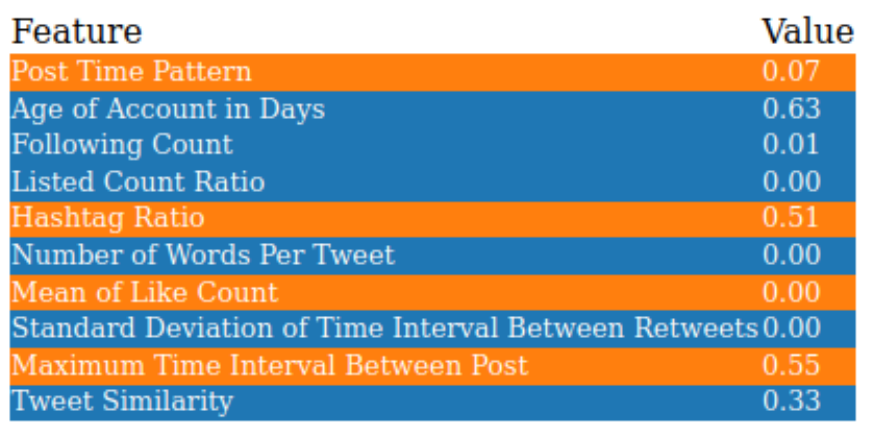}
\end{subfigure}
  \caption{Local Interpretation of misclassified sample using LIME}
  \label{fig:lime_sample_3}
\end{figure*}

Figure~\ref{fig:shap_sample_3} illustrates the local interpretation of incorrectly classified data; in this case, the sample actually represents a human account that was mistakenly identified as an automated account. In this scenario, the model score is significantly impacted by the features indicated by the red color.\\
Figure~\ref{fig:lime_sample_1} depicts the LIME interpretation of a randomly selected automated account. The plot shows the predicted probability for each class, with the bot having the highest probability of 66\%. The significance of each attribute to this prediction is depicted in the center bar plot. The feature emphasized in orange contributes toward the bot class, whereas blue signifies the human class. In this sample, the \texttt{number of post per day} has 15\% feature significance score, followed by the \texttt{account age}, \texttt{number of like count}, and \texttt{number of words per tweet}, and so on. The right plot depicts the top ten features impacting prediction and their associated values. These attributes have a beneficial effect on prediction output.\\
Figure~\ref{fig:lime_sample_2} shows the local interpretation of a human account that was classified correctly. From this figure, we observe that the \texttt{account age} has an 8\% favorable impact on the prediction. Furthermore, the \texttt{mean of like count, mean of reply count}, etc have a positive influence on predicting this sample as human. 
Local interpretation of a misclassified sample is illustrated in Figure~\ref{fig:lime_sample_3}.  From the plot, the account has a 92\% probability of being predicted as human; however, it is actually an automated account. The features marked in blue, such as \texttt{account age, following count, listed count ratio}, and so on, influence the prediction of the XGBoost model towards the human class. Furthermore, the \texttt{account age} is more important, with a 12\% influence score for this prediction. However, features such as \texttt{post time pattern, hashtag ratio},  and \texttt{mean of like count} all contribute to predicting this sample as an automated account. These interpretations make it simple to discover the features driving the performance of the model. 
\section{Summary of Findings}
\label{sec:results}
\subsection{RQ$_1$: Which kinds of IoCs are mainly transmitted through security social media?}
\begin{figure}
    \centering
    \includegraphics[width=0.4\linewidth]{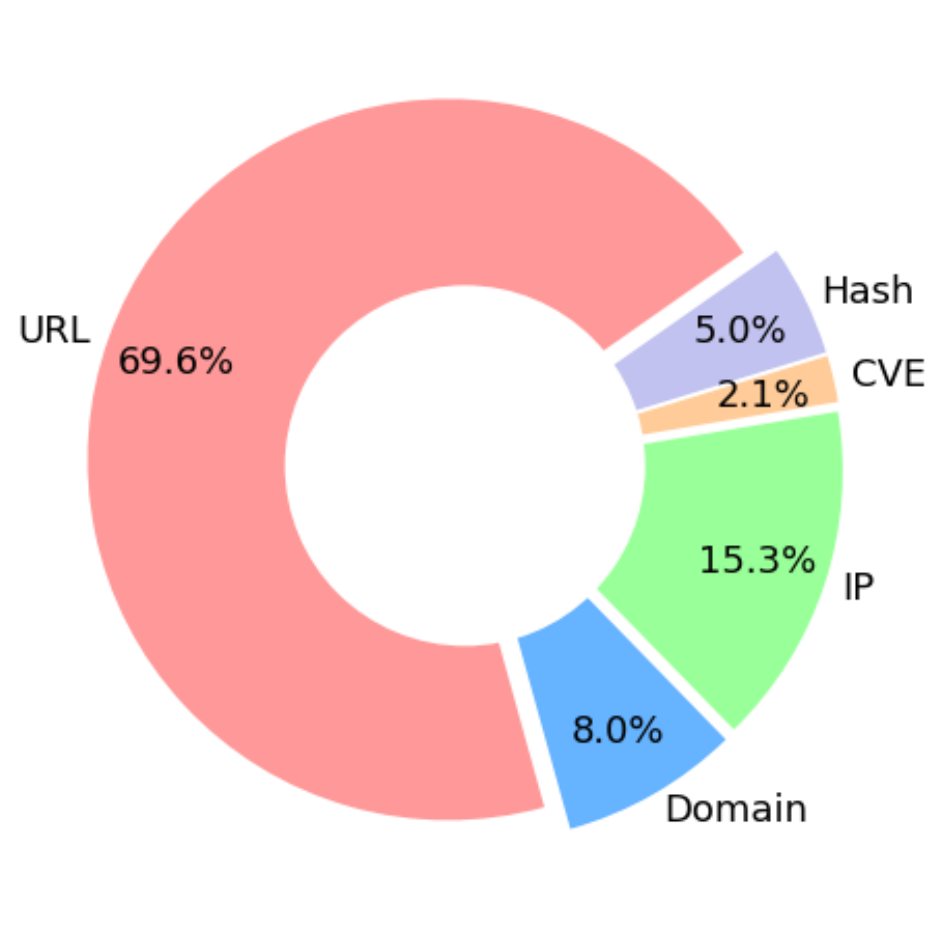}
    \caption{Statistics of IoCs transmitted through security Tweets from January 2021 to July 2021}
    \label{fig:statistic}
\end{figure}

Figure~\ref{fig:statistic} shows the overall statistics of the IoCs reported from January to July 2021. According to these data, we observe that URLs are the most commonly traded IoCs compared to others. From January to July 2021, 69.6\% of URLs were acquired, while only 2.1\% of vulnerability indicators (CVE) were obtained. In general, URLs are the most communicated IoCs through security Tweets.
Additionally, the data we gathered from Twitter provides meaningful information about various types of attacks that happened in the cyber world. On Twitter, most security practitioners reported valuable threat information by the relevant attack keywords/ hashtags. From the relevant tweets that contain IoC, we derived which sort of attacks, such as malware and phishing, are frequently adopted by the attackers. Table~\ref{tab:attack_types} demonstrates the statistics of the most occurring attacks mentioned in the security tweets. From the table, we inferred that 72.7\% of URLS mentioned in the security tweets are associated with phishing attacks. While 93.1\% of file hashes are related to the malware family.
\begin{table}
 \centering
    \caption{The statistics of the most common attack names mentioned in security tweets}
  \footnotesize
    \begin{tabular}{llll} \hline
 \textbf{IoC Type }& \textbf{Phishing} & \textbf{Malware} & \textbf{Others}\\ \hline
  URL & 72.7\% & 4.6\% & 22.7\%\\
  IP & 34.8\%& 50.8\% & 14.4\%\\ 
  Domain & 88.8\% & 3.5\% & 7.7\% \\ 
  File Hashes & 2.7\% & 93.1\%& 4.1\%\\ \hline
  \end{tabular}

  \label{tab:attack_types}
  \end{table}
\subsection{RQ$_2$: To what extent IoCs extracted from social media accounts are reliable?}
We assessed the applicability of Twitter in providing threat intelligence and the reliability of IoCs using metrics such as \textit{correctness, timeliness}, and \textit{overlap}.\\
\textit{\textbf{Correctness: What percentage of IoCs are reported malicious by different threat intelligence services?}}\\
%Correctness is one of the most important aspects to consider when assessing the trustworthiness of IoC.
Correctness is obtained by calculating the number of confirmed IoCs~(threat indicator) reported by at least any threat intelligence services. The percentage of each type of malicious indicator is summarized in Table~\ref{tab:correctness}.
\begin{table*}
\centering
\footnotesize
\caption{The percentage of confirmed IoCs}
\begin{tabular}{lllllll}
\hline
IoC Type  & URL & IP & Domain & Hashes & CVE & Total \\
\hline
IoCs reported on Twitter     & 63313    &   13964 &     7276   &   4545     &   1893 & 90991 \\
Confirmed IoCs reported by any of the TISs &  29743   &   4015  &    4196    &    4458    &    1869 & 44281 \\
\textbf{Percentage of confirmed IoCs}     &   \textbf{46.98}  &  \textbf{28.75}  &      \textbf{57.67}  &    \textbf{98.08}    &   \textbf{98.73} & \textbf{48.67}\\
\hline
\end{tabular}
\label{tab:correctness}
\end{table*}

\begin{figure*}[ht]
    \centering % <-- added
\begin{subfigure}{0.33\textwidth}
  \includegraphics[width=\linewidth]{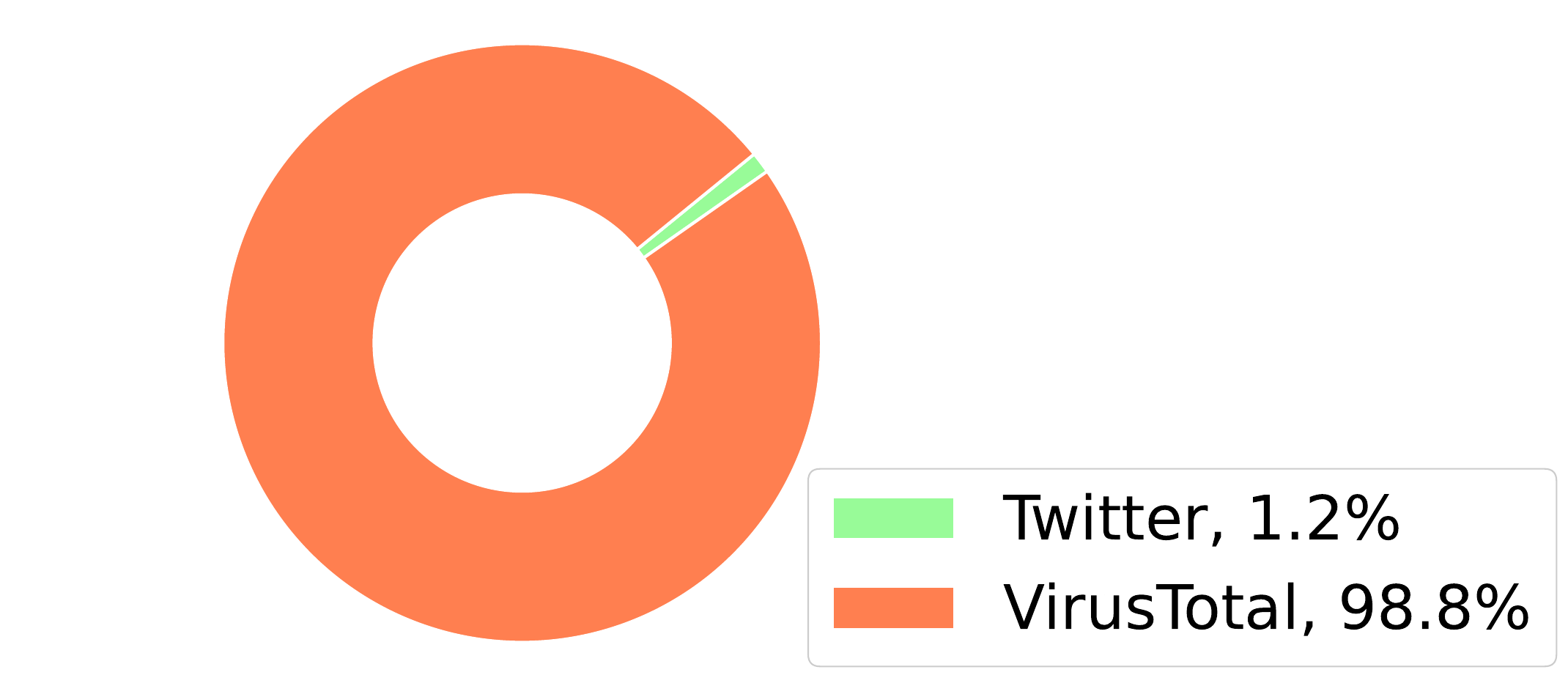}
  \caption{IP}
  \label{fig:timeliness_ip}
\end{subfigure}\hfil % <-- added
\begin{subfigure}{0.33\textwidth}
  \includegraphics[width=\linewidth]{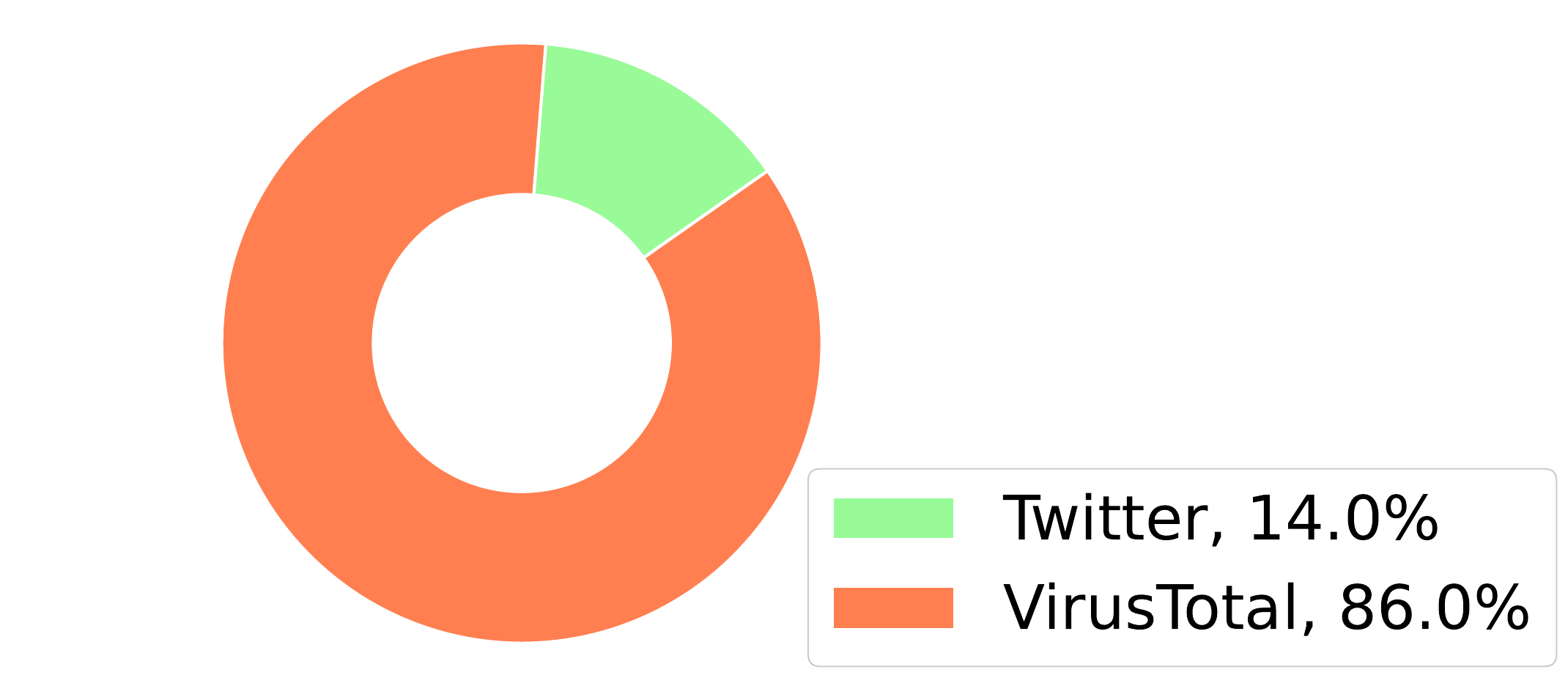}
  \caption{Domain addresses}
  \label{fig:timeliness_domain}
\end{subfigure}
\begin{subfigure}{0.33\textwidth}
  \includegraphics[width=\linewidth]{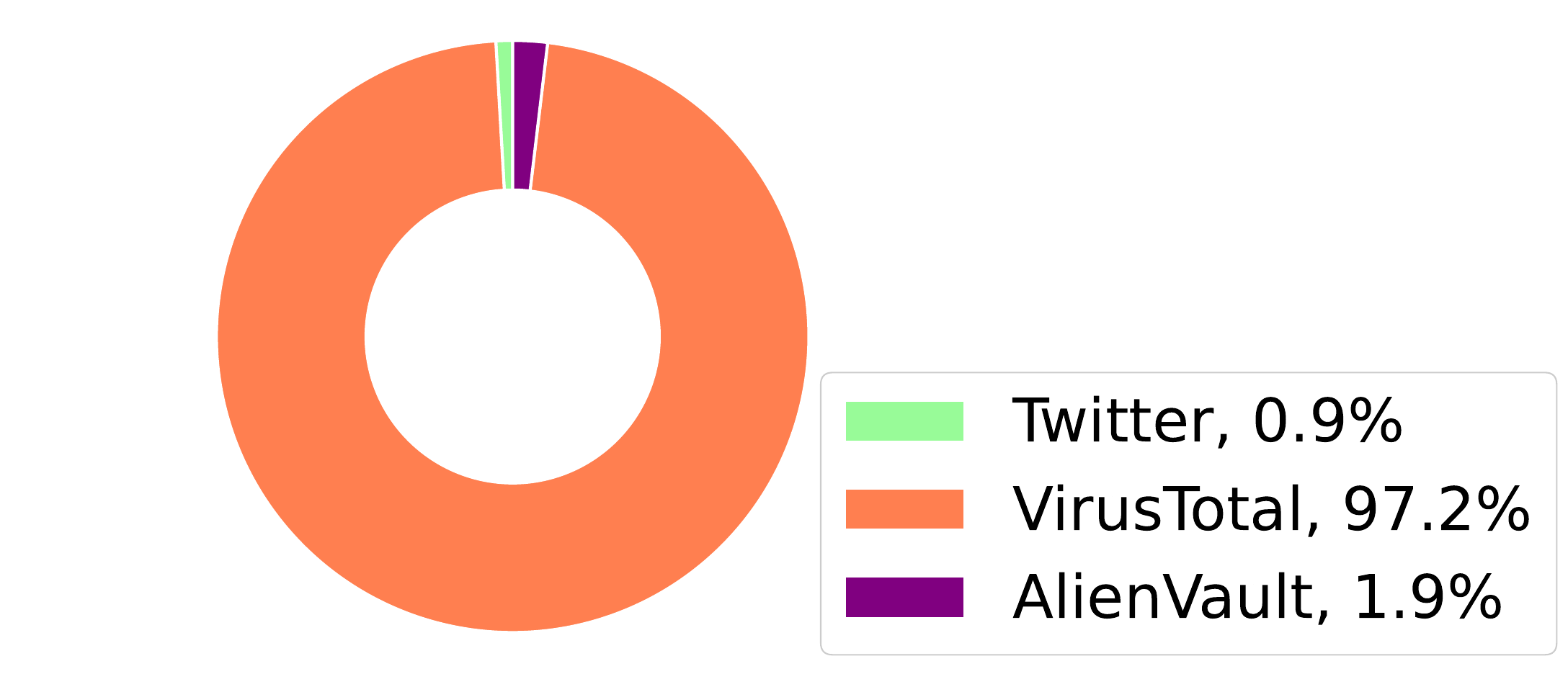}
  \caption{File Hashes}
  \label{fig:timeliness_hash}
\end{subfigure}\hfil % <-- added
\begin{subfigure}{0.33\textwidth}
  \includegraphics[width=\linewidth]{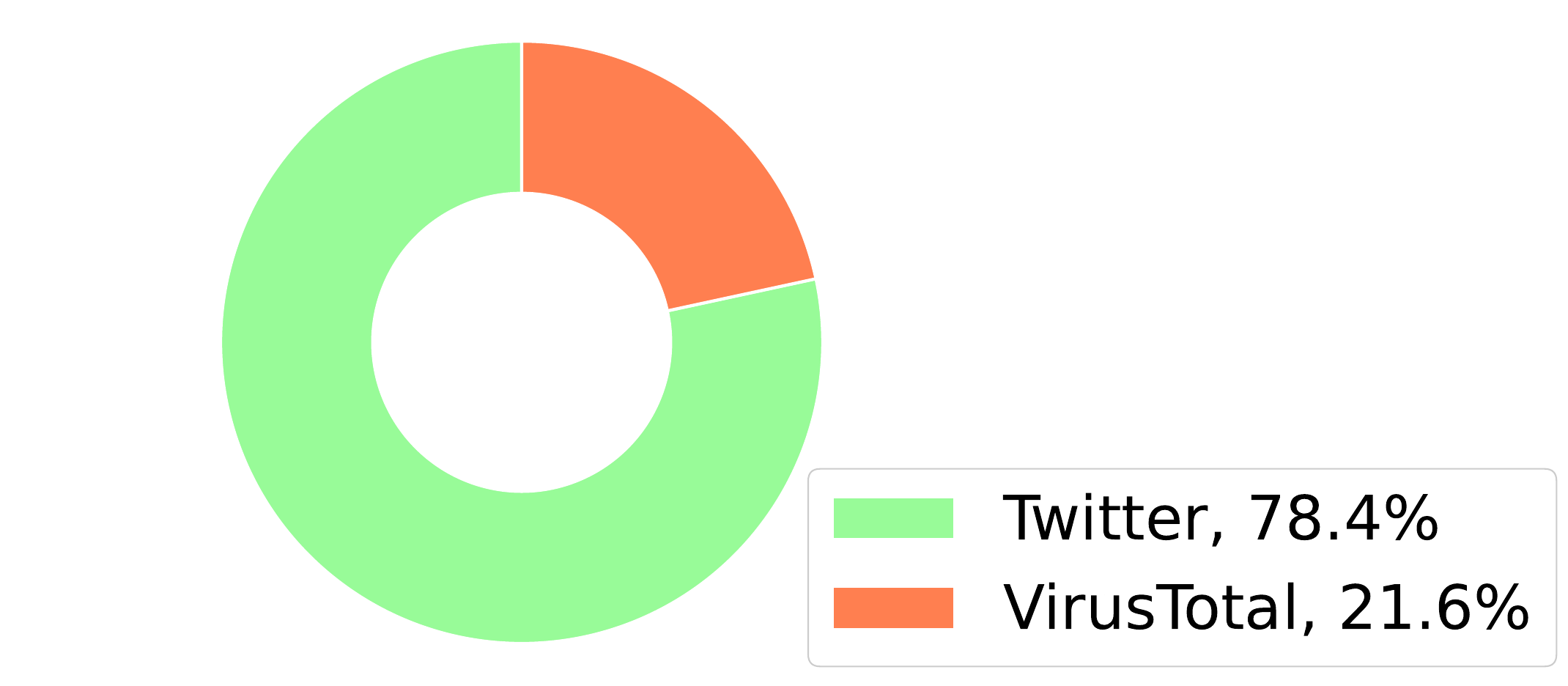}
  \caption{URL}
  \label{fig:timeliness_url}
\end{subfigure}
\begin{subfigure}{0.33\textwidth}
  \includegraphics[width=\linewidth]{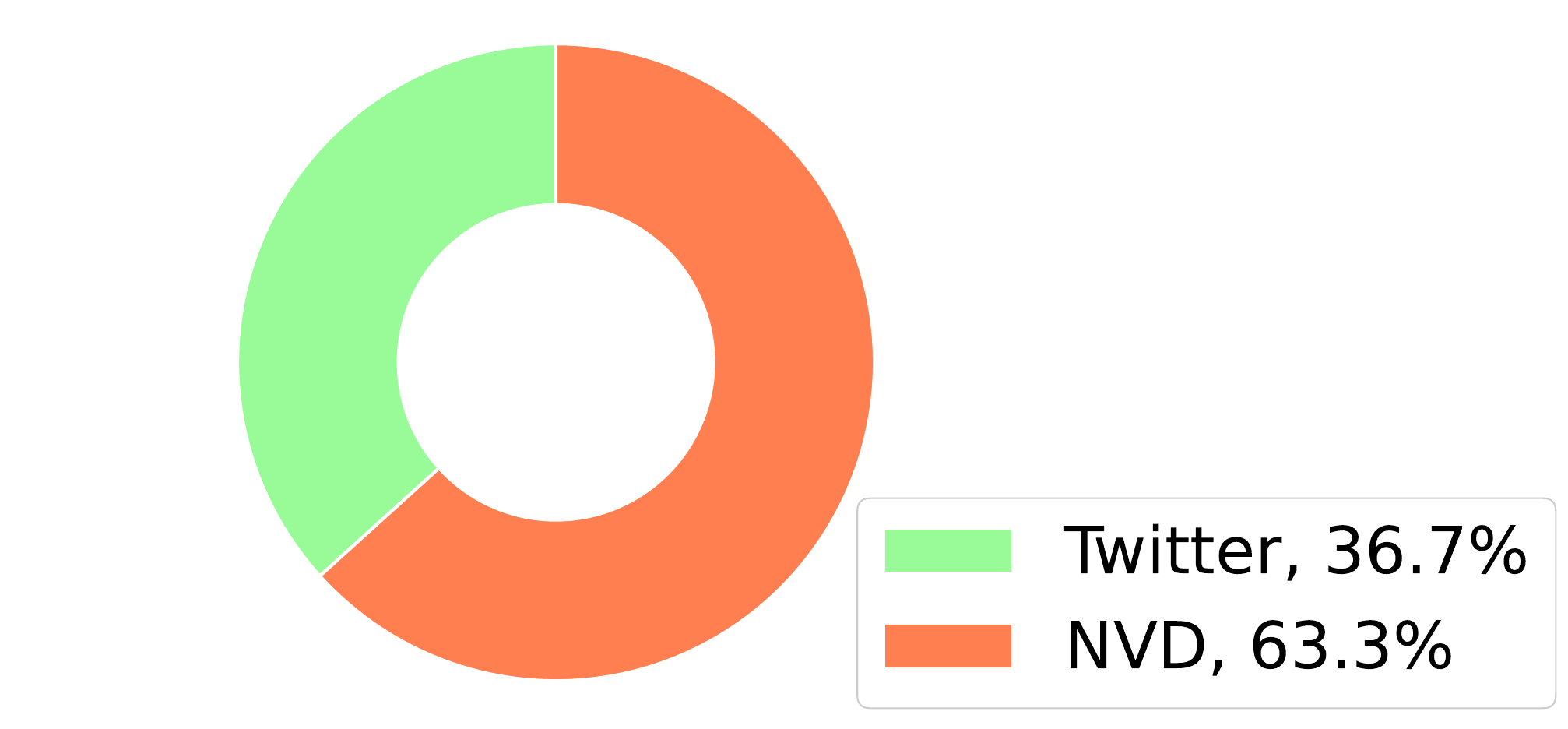}
  \caption{CVE}
  \label{fig:timeliness_cve}
\end{subfigure}\hfil % <-- added
\caption{Comparison of Reported Date of IoCs with various Threat Intelligence Services}
\label{fig:timeliness}
\end{figure*}

\begin{figure*}[ht]
    \centering % <-- added
\begin{subfigure}{0.33\textwidth}
  \includegraphics[width=\linewidth]{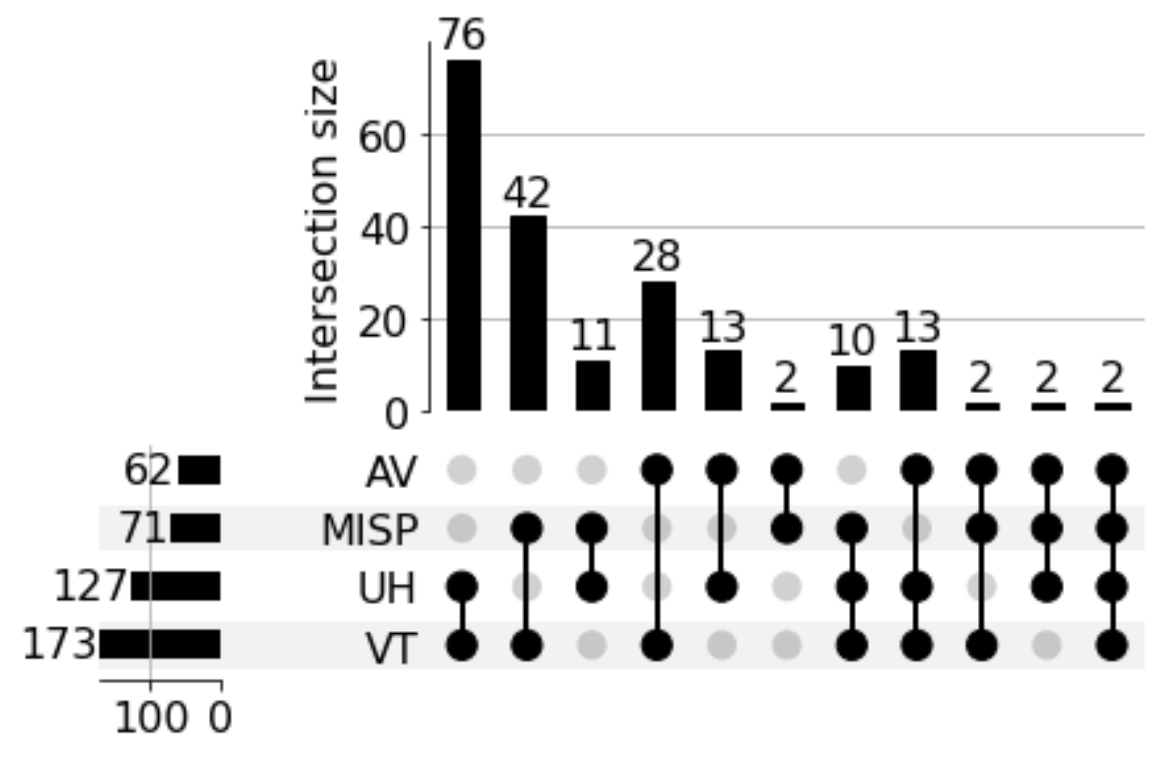}
  \caption{URL}
  \label{fig:overlap_url}
\end{subfigure}\hfil % <-- added
\begin{subfigure}{0.33\textwidth}
  \includegraphics[width=\linewidth]{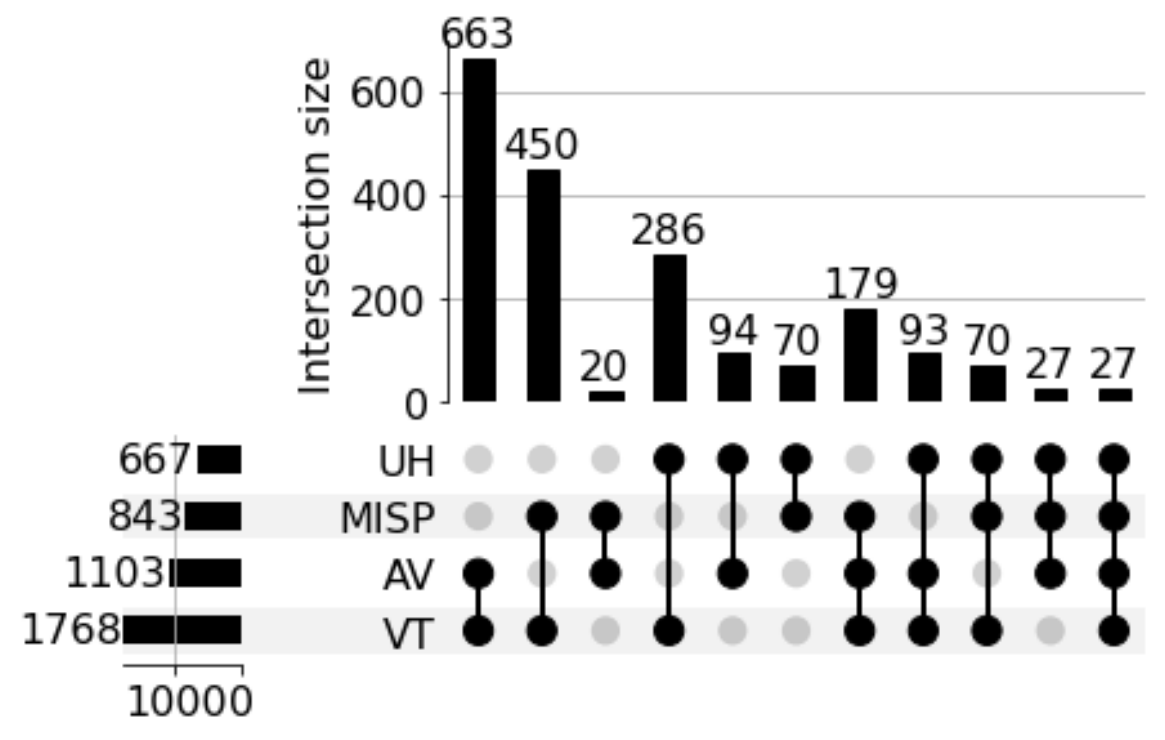}
  \caption{IP}
  \label{fig:overlap_ip}
\end{subfigure}
\begin{subfigure}{0.33\textwidth}
  \includegraphics[width=\linewidth]{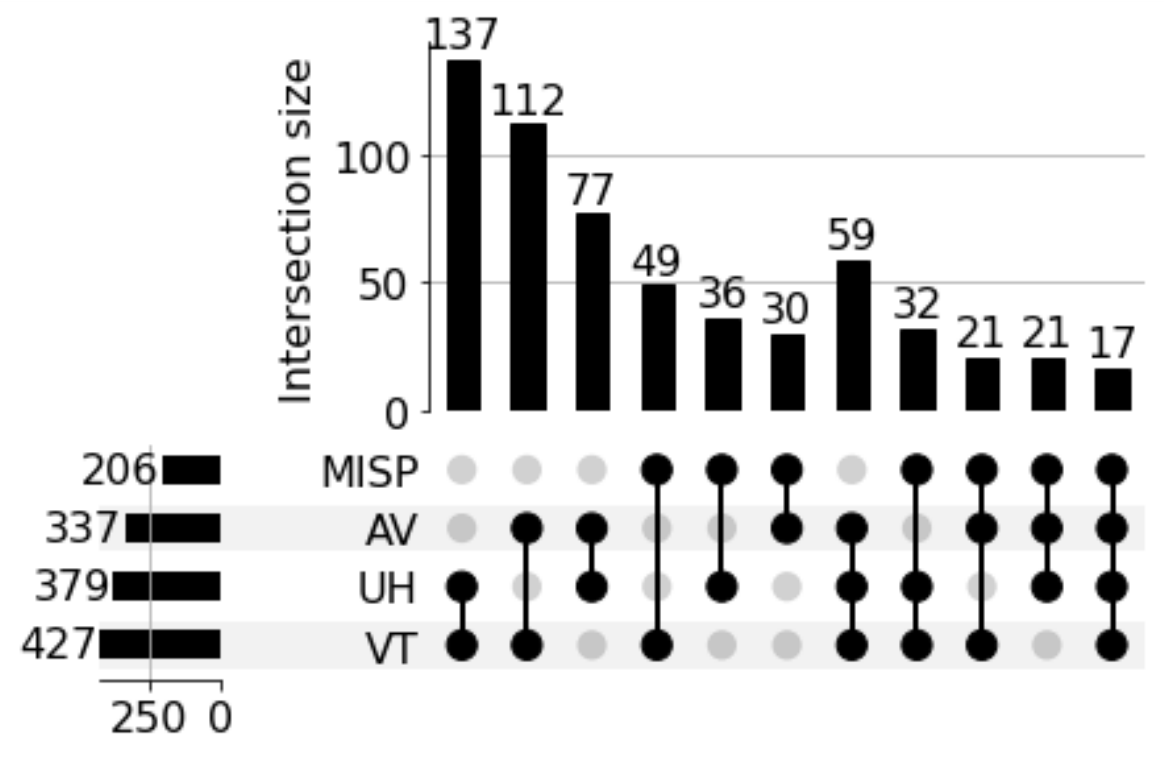}
  \caption{Domain}
  \label{fig:overlap_domain}
\end{subfigure}\hfil % <-- added
\begin{subfigure}{0.7\textwidth}
  \includegraphics[width=\linewidth]{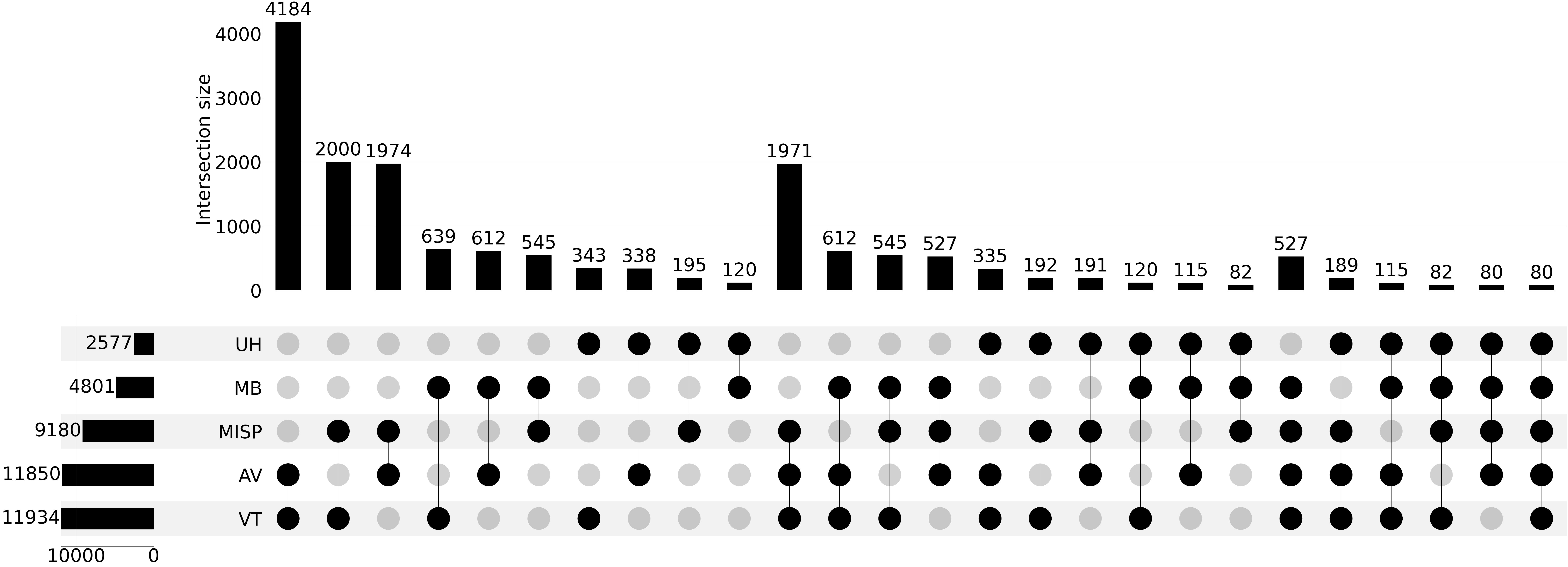}
  \caption{Hash}
  \label{fig:overlap_hash}
\end{subfigure}
\begin{subfigure}{0.25\textwidth}
  \includegraphics[width=\linewidth]{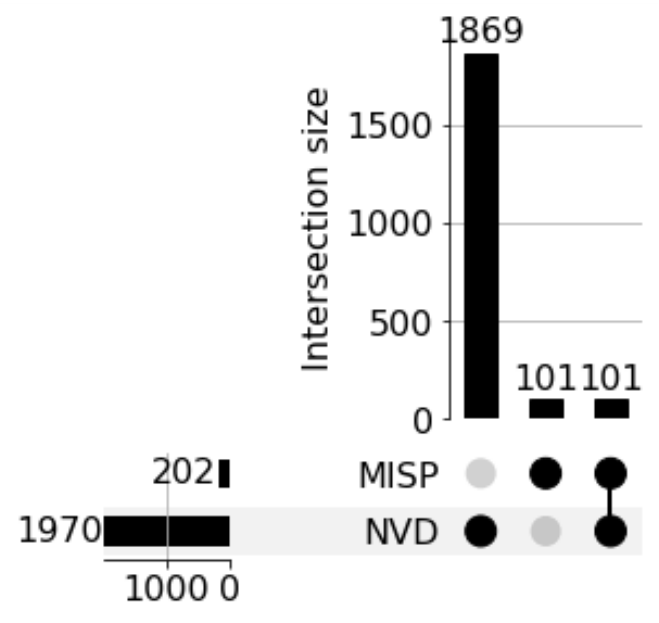}
  \caption{CVE}
  \label{fig:overlap_cve}
\end{subfigure}
\caption{The number of common IoCs among various Threat Intelligence Services}
\label{fig:overlap}
\end{figure*}

Our investigation using threat intelligence platforms shows that 48.67\% of IoCs reported on Twitter are actually malicious. Further, 98.08\% of hashes reported in tweets constitute a threat signal. Recently, social media has been increasingly used to report signs of vulnerability. 98.73\% of CVEs mentioned in tweets are already in NVD; however, 1.27\% of novel vulnerabilities are released by Twitter users and not yet in the national vulnerability database. Furthermore, 46.98\% of URLs and 56.7\% of domains cited by security users are cyber forensic evidence associated with a threat or attack. Regarding IP addresses, 28.75\% of IP addresses related to threats stated in tweets are also in various threat intelligence services.\\
\textit{\textbf{Timeliness: How fast does social media share its IoCs compared to other threat intelligence services?}}\\
The timeliness metric is a key performance indicator that determines which platforms disclose IoCs as early as possible. We determined the timeliness metric by comparing the published date of the IoC on Twitter and the initial submission date of the IoC on various threat intelligence services. As TIS do not provide analysis dates for certain confirmed IoCs, we were only able to compare 39,402 out of the total 44,281 IoCs. The percentage of IoCs reported earlier by Twitter and various threat intelligence services are plotted in Fig~\ref{fig:timeliness}. For the IP, domain addresses, and URLs, we compared the timeliness with only VirusTotal since Table~\ref{tab:IoC_Status} clearly shows that VirusTotal is the TIS delivering the highest number of indicators. Other intelligence services identified a few of the indicators as malevolent. So we considered VirusTotal as the baseline platform, which scans all sorts of IoCs and conveniently delivers the first reported date for almost all IoCs, with the exception of a few IP addresses. Figure~\ref{fig:timeliness_url} indicates that Twitter delivers timely threat warnings in the form of URLs by reporting 78.4\% of URLs earlier than VirusTotal. In the case of IP addresses and domains, 1.2\% of malicious IPs are reported earlier by Twitter, and 14\% of malicious domain addresses are timely reported by Twitter. Although the proportion of timeliness for IP and domain addresses is lower, Twitter still contributes with a few confirmed IoCs ahead of the VirusTotal. We observed VirusTotal and AlienVault for file hashes since both platforms reported almost all hashes as malicious indicators. In terms of timeliness regarding hashes, VirusTotal outperforms Twitter and AlienVault with a 97.2\% of early warning. We estimated the timeliness of CVEs between Twitter and NVD since NVD is the primary vulnerability database based on and completely synced with the CVE records; all changes to CVE appear instantly in NVD. The investigation shows that social media like Twitter report 36.7\% of CVEs earlier than NVD. Through our analysis, we found that Twitter provided 24,479 IoCs before TIS. Therefore, it is evident from this that Twitter offers reliable and timely warnings against novel threats and vulnerabilities.
\\
\textit{\textbf{Overlap: How many IoCs in social media exist in other Threat Intelligence Services?}}\\
Determining the popular IoC is a crucial task for a threat to community and security practitioners. If we can find out which artifacts are currently involved in malicious activities, it is easier for security analysts and researchers to take essential steps to prevent further threat activities. We discovered the important IoCs by finding the intersection of the malicious obtained from the various threat intelligence services. Since not all threat intelligence systems can scan all types of IoCs, we determined the intersection for each indicator. The upset plot shown in Figure~\ref{fig:overlap} is used to illustrate the findings about the overlap. Figure~\ref{fig:overlap_url} shows that URLs can be scanned on four different platforms, including VirusTotal, AlienVault, UrlHaus, and MISP. Two URLs are common among all four of these platforms.
According to the total malicious URL count, the number of common URLs is minimal~(say two), which is due to the number of malicious URLs reported by AlienVault and MISP.  AlienVault and MISP reported very few URLs as malicious. If we exclude AlienVault and MISP, 76 URLs are common among VirusTotal and UrlHaus. Similar to the URL, for IP and domain addresses, we computed the intersection between the VirusTotal, AlienVault, UrlHaus, and MISP. The most pertinent IoCs, in terms of IP and domain addresses, are 27 and 17, respectively. There are 663 IP addresses common among VirusTotal and AlienVault, as well as 137 domain addresses common among VirusTotal and UrlHaus. With the exception of NVD, all threat intelligence services allow us to examine the file hashes, so we have taken the intersection between them. The most important evidence of compromise, according to our investigation, is 80 file hashes. Additionally, we can see from Figure~\ref{fig:overlap_hash} that almost all hashes found in VirusTotal and AlienVault are involved with threat activities.
Although NVD is the primary database of vulnerability indicators in the case of CVE, we can also analyze CVE in MISP. However, MISP only exposes us to 101 CVEs out of the 1970 CVES we collected, which results in the most common vulnerability.
\subsection{RQ$_3$: To what proportion are the automated accounts delivering IoCs through social media?}
We employed the XGBoost model to determine whether the IoCs posted on Twitter originated from human-maintained or automated accounts. Delving into the specifics of IoC activity, the dataset contains a notable 90,991 IoCs after the removal of retweets and duplicates. Of these IoCs, a total of 1,749 accounts actively participated in posting them. Interestingly, human accounts were responsible for a significant majority of these postings, contributing a total of 77,892  IoCs and involving 1,717 individual human accounts. In contrast, automated accounts posted 13,099 IoCs, with their participation limited to 32 accounts. It is worth noting that, despite the limited number of automated accounts (i.e., 32), they were identified to be responsible for sharing about 14.4\% of the overall IoCs in our dataset. This emphasizes the potential role that each bot account can have for successfully spreading large portions of cyber threat information.
Moreover, we analyzed the percentage of Twitter accounts sharing 44,281 confirmed IoCs. 
These confirmed IoCs were distributed among 817 accounts engaged in their dissemination. Although human accounts played the primary role in the dissemination of confirmed IoCs, sharing 38,149 confirmed IoCs across 798 distinct human accounts, automated accounts also contributed to the propagation of confirmed IoCs, with 6,132~(13.8\%) IoCs attributed to 19 bot accounts. 
Notably, the percentages of confirmed IoCs over total shared IoCs are similar for the two types of accounts (i.e., 48.98\% for human-operated accounts and 46.81\% for automated accounts), suggesting that an IoC maintains an almost consistent level of reliability irrespective of whether it is disseminated by an automated or human-operated account.  
Additionally, we analyzed the prevalence of automated accounts sharing confirmed IoCs before TISs and found that they initiated 186 cyber artifacts. To summarise, our analysis confirms that automated accounts are also involved and play a complementary role in disseminating confirmed IoCs on Twitter.
\section{Discussion}
\label{sec:discussion}
Although our study involved collecting Twitter data over seven months, it is worth emphasizing that our dataset is quite substantial, consisting of an impressive 1.39 million English tweets. This extensive dataset empowered us to conduct a thorough and comprehensive analysis, thereby yielding valuable insights into our research questions. While we acquired Twitter data for IoC sources in our study, it is likely that we also gathered non-security tweets or security news devoid of IoCs. To partially mitigate this issue, like previous researchers~\cite{kristiansen2020cti,behzadan2018corpus,shin2021twiti}, we used a set of hashtags and/or keywords. However, the tweets might not be relevant because they can include blog URLs and tutorial discussions with the same hashtags. We developed an automated system with the help of a deep learning model to identify relevant tweets bearing IoCs. Furthermore, to limit the irrelevant IoCs, we purged certain unnecessary URLs like Tweet URLs, Facebook and Youtube URLs, etc., using regular expression rules. According to Table~\ref{tab:correctness}, URL is the most prevalent IoC sent via security tweets. It might be because malicious URLs are frequently used as the centerpiece of several cyber incidents, including malware, phishing, and spamming\footnote{https://resources.infosecinstitute.com/topic/threat-hunting-for-urls-as-an-ioc/}. Additionally, it is simple for attackers to deliver the URLs via the existing communication systems. Moreover, the search criteria used to acquire the data for this study included terms like ``phishing," ``malware," ``0day," ``botnet," ``spyware," ``ransomware," and others. However, as shown in Table~\ref{tab:IoC_Status}, the percentage of suspicious IPs identified as malicious is relatively low across all TIS. Since attackers frequently change IP addresses~\cite{villalon2022key}, it is possible that threat intelligence services rarely examine and store IP addresses in their feeds. Thus according to our findings, IP addresses are a poor indicator of threat activities.
\par
Table~\ref{tab:IoC_Ststistics} illustrates the malicious status of IoC in various TIS, which reflects how well threat intelligence services can deliver threat information. We deduced from the table that, with the exception of CVE, VirusTotal could scan all forms of IoCs and offer a reasonable outcome. So, based on our study and previous studies~\cite{shin2021twiti}, we recognized Virustotal as a core intelligence service. MISP has recently been adopted by many security experts, although the volume of events and threat information recorded in MISP remains relatively low. In the future, we plan to integrate the MISP by adding the indicators we have identified as malicious during our research into MISP so that we can discover more about indicators and the events associated with them. The automation system for validating the source of the security tweets achieved 0.814 macro and 0.925 weighted F1 scores, but it provides some false positives. Since automated accounts and advertising agencies have similar traits, such as the number of followers and retweets, the system seldom makes classification errors. The false positives can be minimized by continuous learning~\cite{chen2023continuous} and expanding the dataset.
\section{Conclusions and Future Work}
\label{sec:conclusions}
To address Advanced Persistent Threats (APTs) effectively, security firms increasingly rely on CTI, which offers valuable insights into attackers' motives and helps predict future threats. Nowadays, social media has emerged as a crucial platform for sharing threat intelligence. In this study, we gathered cyber-related posts from social media and extracted relevant Indicators of Compromise (IoCs) using regular expressions. To filter out irrelevant posts, we employed a binary classifier with an impressive F1-score of 98.80\% and a detection rate of 99.65\%. Utilizing Threat Intelligence Systems (TIS), we assessed the reliability of IoCs based on metrics like correctness, timeliness, and overlap. Our analysis revealed that URLs are the most commonly shared IoCs on social media. According to the correctness metric, 48.67\% of these IoCs were genuine threat indicators. In terms of timeliness, Virustotal typically provided early warnings for IoC-related threats, while social media excelled in delivering early alerts specifically for URLs. Through the overlap metric, we identified critical artifacts currently involved in malicious activities. We also assessed the role of automated accounts in spreading IoCs, employing various machine learning and deep learning models to differentiate between human and automated sources. XGBoost demonstrated the best performance, with a macro F1-score of 0.814 and a weighted F1-score of 0.925. Furthermore, we demonstrated that both human-operated and automated accounts disseminate IoCs. These findings suggest that social media is becoming increasingly popular for delivering relevant and accurate suspicious IoCs. In the future, we plan to extract detailed information about a threat, such as malware name, attacker, attack type, target, and similar, from the security blogs. We also plan to investigate the role of technical blogs in delivering threat intelligence. Furthermore, we plan to incorporate additional metrics to enhance the evaluation of IoC reliability.
\bibliographystyle{unsrt}
\bibliography{bibilio} 
\end{document}